\documentclass[sn-mathphys-num]{sn-jnl}

\usepackage{graphicx}%
\usepackage{multirow}%
\usepackage{amsmath,amssymb,amsfonts}%
\usepackage{amsthm}%
\usepackage{mathrsfs}%
\usepackage[title]{appendix}%
\usepackage[rgb,dvipsnames]{xcolor}%
\usepackage{textcomp}%
\usepackage{manyfoot}%
\usepackage{booktabs}%
\usepackage{algorithm}%
\usepackage{algorithmicx}%
\usepackage{algpseudocode}%
\usepackage{listings}%
\usepackage{indentfirst}
\usepackage{makecell}
\usepackage{tabularx}
\usepackage{caption}
\usepackage{subcaption}
\usepackage{enumitem}
\usepackage[most]{tcolorbox}
\usepackage{anyfontsize}
\usepackage{adjustbox}
\usepackage{hyperref}

\begin{document}

\title[Article Title]{Harnessing Large Language Models for Virtual Reality Exploration Testing: A Case Study}

\author[1]{\fnm{Zhenyu} \sur{Qi}}\email{qzydustin@arizona.edu}

\author[1]{\fnm{Haotang} \sur{Li}}\email{haotangl@arizona.edu}

\author[2]{\fnm{Hao} \sur{Qin}}\email{hqin@arizona.edu}

\author[3]{\fnm{Kebin} \sur{Peng}}\email{pengk24@ecu.edu}

\author[1]{\fnm{Sen} \sur{He}}\email{senhe@arizona.edu}

\author*[4]{\fnm{Xue} \sur{Qin}}\email{xue.qin@villanova.edu}

\affil[1]{\orgdiv{Department of Electrical and Computer Engineering}, \orgname{The University of Arizona}}

\affil[2]{\orgdiv{Department of Mathematics}, \orgname{The University of Arizona}}

\affil[3]{\orgdiv{Department of Computer Science}, \orgname{East Carolina University}}

\affil*[4]{\orgdiv{Department of Computing Sciences}, \orgname{Villanova University}}

\abstract{As the Virtual Reality (VR) industry expands, the need for automated GUI testing is growing rapidly. Large Language Models (LLMs), capable of retaining information long-term and analyzing both visual and textual data, are emerging as a potential key to deciphering the complexities of VR's evolving user interfaces. In this paper, we conduct a case study to investigate the capability of using LLMs, particularly GPT-4o, for field of view (FOV) analysis in VR exploration testing. 
Specifically, we validate that LLMs can identify test entities in FOVs and that prompt engineering can effectively enhance the accuracy of test entity identification from $41.67\%$ to $71.30\%$. Our study also shows that LLMs can accurately describe identified entities' features with at least a $90\%$ accuracy rate.
We further find out that the core features that effectively represent an entity are color, placement, and shape.
Furthermore, the combination of the three features can especially be used to improve the accuracy of determining identical entities in multiple FOVs with the highest F1-score of $0.70$. 
Additionally, our study demonstrates that LLMs are capable of scene recognition and spatial understanding in VR with precisely designed structured prompts. Finally, we find that LLMs fail to label the identified test entities, and we discuss potential solutions as future research directions.} 


\maketitle

\section{Introduction} \label{sec:intro}

The virtual reality (VR) industry is rapidly expanding, with the global market expected to reach approximately 32.64 billion USD by the end of 2024~\cite{VR_market}. 
Due to its ability to provide immersive and simulated experiences, VR technology is increasingly being utilized in various sectors, including gaming, healthcare, and education. 
As a result, thorough testing of VR graphical user interfaces (GUIs) is essential to ensure the quality of user interactions and application functionality. 
Existing research on VR GUI testing encompasses areas such as the development of testing frameworks~\cite{vrtest, DyTRec}, camera coverage analysis~\cite{vrguide}, studies on code practices~\cite{rzig2023vr_testing}, and user simulation testing~\cite{SIM2VR}. 
However, none of these approaches directly address the automation of VR GUI exploration testing. The closest related techniques are focused on 3D games~\cite{Angelo2023ASE, Joseph2022ATEST}, where new frameworks are proposed to improve testing strategies like search testing.
However, the fragmented nature of the VR market, which includes more than six major brands and five development engines, makes it challenging to adapt these techniques to the majority of VR applications.

GUI testing on non-VR applications also faces similar challenges in various software and hardware configurations.
One popular solution is to design a black-box GUI 
automation~\cite{yShengcheng} testing based on image processing~\cite{Jieshan, owleye} and large language models (LLMs)~\cite{DroidAgent}.
Recent LLMs, such as GPT-4o~\cite{openai2024gpt4o} and BERT~\cite{BERT}, show their ability to process and generate data in mixed formats, including text, audio, images, and video. This capability has spurred the development of numerous LLM-based GUI automation techniques~\cite{zhang2024largelanguagemodelbrainedgui}, spanning both general testing frameworks~\cite{GPTDroid, AUITestAgent, AXNav} and task-specific testing approaches~\cite{QTypist, CrashTranslator}. LLMs are also widely seen in building virtual assistants due to their ability to understand app interfaces and establish conversations with users.
One paper~\cite{qin2024utilizing} has discussed the potential of LLM in completing the VR exploration tasks, but the scale is very small and only focuses on a limited number of tasks.

In this paper, we construct a large-scale dataset of $270$ screenshots with labeled groundtruth and explore the potential of using LLMs, specifically GPT-4o, to assist VR exploration testing.
To understand how LLM could assist in GUI exploration testing, we first review the core procedures involved, which are
1) identify the test entity, 2) navigate to the entity, 3) simulate interaction, and 4) verify the response.
As accurate entity identification and subsequent navigation are prerequisites for further steps, we focus our study on these first two steps. 
Therefore, we design the following five research questions to measure LLM's capability and efficiency by analyzing the user's field of view (FOV) in the VR app.

\noindent Our main research questions are: 
\begin{itemize}[]
    \item \textbf{RQ1}: Can prompt engineering be used to enhance the accuracy of entity detection from a FOV in VR?\\
    \textbf{Motivation.}
    The initial question in our study aims to validate the LLM's capability of detecting entities in an arbitrary FOV, and to leverage prompt engineering for enhanced accuracy.
    \\
    \textbf{Answer.} 
    We find the LLM, GPT-4o, can detect entities on given FOVs with an average accuracy of $41.67\%$.
    After adopting prompt engineering, specifically Chain of Thought Prompting, we achieve an improved accuracy of $71.30\%$.

    \item \textbf{RQ2}: What are the core features for identifying test entities in FOV, and how accurately can LLM describe them?\\
    \textbf{Motivation.} 
    After detecting the entities, the LLM should be able to describe them in a way that a human tester can understand.
    In this study, we employ the optimal prompt from RQ1 for entity identification. We then ask LLM to describe the features of the detected entities.
    Finally, we evaluate their accuracy and examine their significance in representing the entity.
    \\
    \textbf{Answer.}
    We collect nine cumulative features from the literature and then contextualize them to five to fit 
    our research goals. Among the five features, we identify color, placement, and shape as three core
    features that GPT-4o uses to identify entities. The description accuracies are $94.80\%$, $95.45\%$, and $96.10\%$, respectively.

    \item \textbf{RQ3}: How accurate is LLM in scene~\cite{def_scene} recognition and spatial understanding through FOV analysis?\\
    \textbf{Motivation.}
    After exploring individual entities from a single FOV, we want to measure the LLM's visual and spatial
    understanding of the scene through FOV.   
    This spatial understanding is crucial for designing future exploration testing, particularly when users navigate to test entities in 3D space. As users move, recognizing the scene from partial FOV views is very important. \\
    \textbf{Answer.}
    In this study, we ask GPT-4o to describe the scene in a few words, e.g., bathroom, dining room, or living room.
    Later, we ask GPT-4o to structurally describe the spatial relationship between the user and the target entity, and then 
    respond with a structured answer containing three dimensions: horizontal, vertical, and depth. 
    Ultimately, we assess the answers and verify their correctness.
    The evaluation shows that $83.12\%$ of scenes can be correctly recognized, and $92.86\%$ of spatial relationships can be correctly understood. 
    
    \item \textbf{RQ4}: Can LLM be used to label the identified entities in VR?\\
    \textbf{Motivation.}
    During exploration testing, after finalizing the target entity and before interacting with it, the next step is to locate or access it using the VR controls.
    In VR devices, reaching or selecting a GUI entity, such as a chair in the room or a cup on the table, is done by moving the joystick so that its raycast line overlaps or points to the target entity.
    A VR application usually uses the coordinates to decide such overlaps.
    In this research question, we want to test the LLM's capability to visually label the detected entities. 
    Then, the visual label information, such as coordinates, can be converted to control signals in the future to help access the GUI entities.
    \\
    \textbf{Answer.}
    In this study, we propose two sets of approaches for using GPT-4o to generate visual labels to comprehensively 
    measure the capability of GPT-4o in this image-processing task. 
    The first set involves directly asking GPT-4o to draw bounding boxes around entities in the raw/compressed FOV images. 
    The second set collects the coordinates of the bounding boxes from GPT-4o, and then draws bounding boxes locally based on the coordinates.
    The qualitative results show that GPT-4o, due to the black-box compression issue, is not capable of accurately labeling the identified entities in VR.
    
    \item \textbf{RQ5}: Can core features enhance the accuracy of determining the same entities in multiple FOVs?\\
    \textbf{Motivation.}
    Once the user starts to move around and explore entities from different directions, it is important to track the interactions and record them correctly. 
    Understanding whether an entity is the same or not is essential for calculating the testing coverage.
    In the last question, we aim to evaluate whether LLM could support such determinations and how accurate they are.
    We are also interested in evaluating whether the detected core features could improve the accuracy of the determinations.
    \\
    \textbf{Answer.}
    The study results show that without considering any features, the F1-score of the baseline is $0.63$.
    Among all the core features, the combination of color, shape, and placement
    achieves the highest F1-score of $0.70$, with the highest Precision of $0.67$ and the highest Recall of $0.74$.

\end{itemize}

\noindent The contributions of this paper include:
\begin{itemize}
    \item Conduct the first quantitative and qualitative case studies to explore LLMs in VR exploration testing, addressing five key research questions.
    \item Create an open-source dataset including VR screenshots and labeled groundtruth.
    \item Design an efficient prompt for entity detection and feature identification in VR applications from arbitrary FOV images.
    \item Provide a detailed analysis of LLM effectiveness for VR tasks and discuss potential solutions for identified limitations.
\end{itemize}

The rest of the paper is organized as follows:
Section~\ref{sec:data_construction} discusses the dataset construction.
Section~\ref{sec:method} presents the study's methodology, detailing the five study designs.
Section~\ref{sec:evaluation} describes the evaluation that answers all five research questions, followed by the observed limitations discussion in Section~\ref{sec:discussion}, and the threats to validity in Section~\ref{sec:thread}.
Section~\ref{sec:related} includes related work. And we conclude our research in Section~\ref{sec:conclusion}.

\section{Dataset Construction} \label{sec:data_construction}

Our research goal and experimental design prioritize virtual environments that closely resemble real-world settings. This is because the study measures LLM's capability and performance in black-box VR exploration testing, which works best with general common knowledge.
The most recent and most extensive VR screenshot dataset from real VR apps is created by Shuqing et al.~\cite{3660803}.
However, most of the images from this dataset are from fantasy scenes with surreal designs and lighting; thus, they do not align well with our experimental needs. Moreover, in RQ4 and RQ5, we also need a sequence of diverse perspectives from the same scene, which is absent in this dataset. 
Therefore, we construct a new dataset\footnote{The dataset is available at 
\url{https://github.com/qzydustin/LLM4VR}} of VR screenshots to provide sufficient data for evaluating the LLM and addressing all five research questions. 

\subsection{Field of View Creation}
\label{subsec:fov_creation}
\begin{figure}[!b] 
    \centering
    \begin{subfigure}{0.32\textwidth}
        \includegraphics[width=\linewidth]{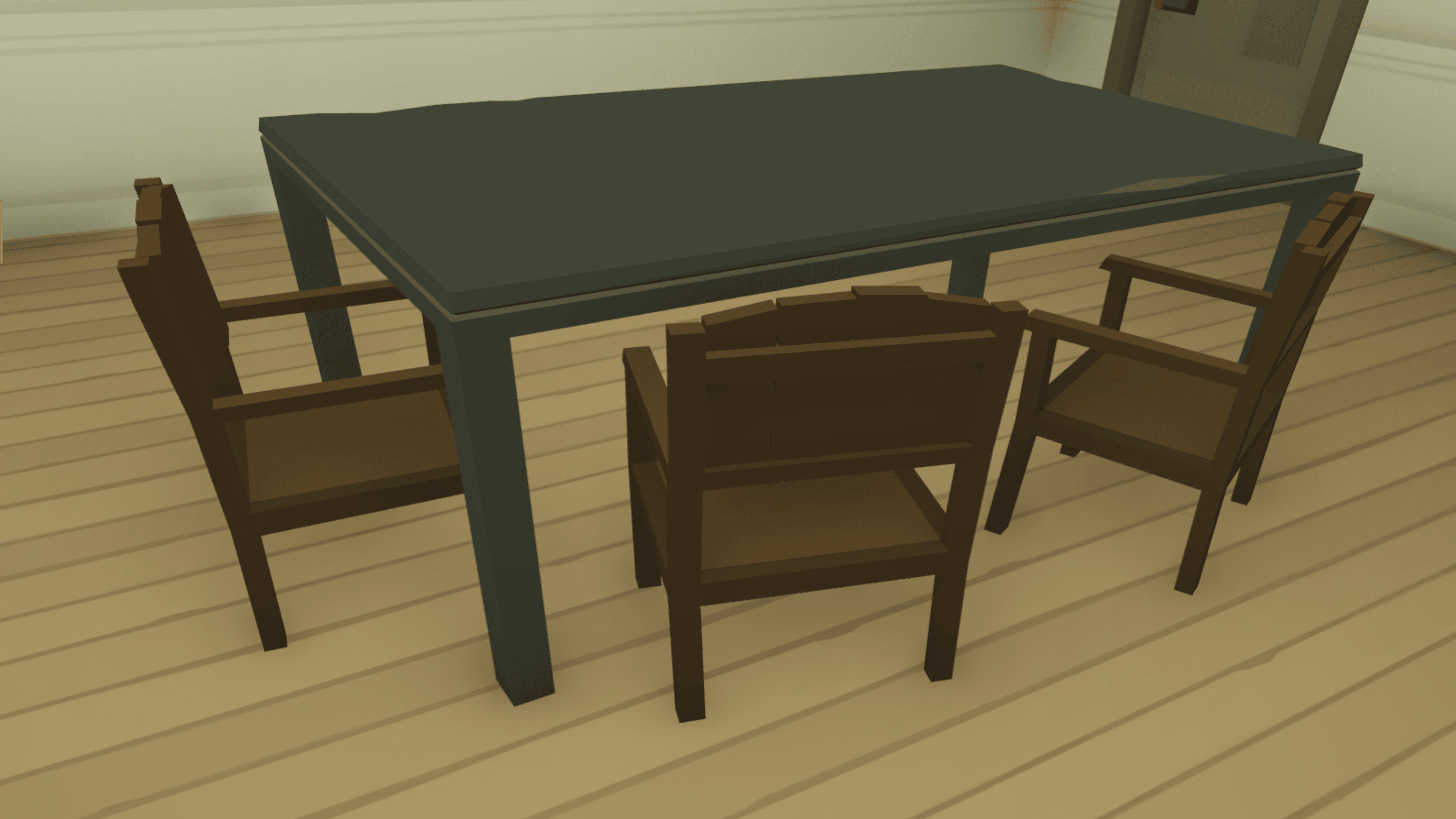}
        \caption{Easy, Light On, Front}
    \end{subfigure}
    \hfill
    \begin{subfigure}{0.32\textwidth}
        \includegraphics[width=\linewidth]{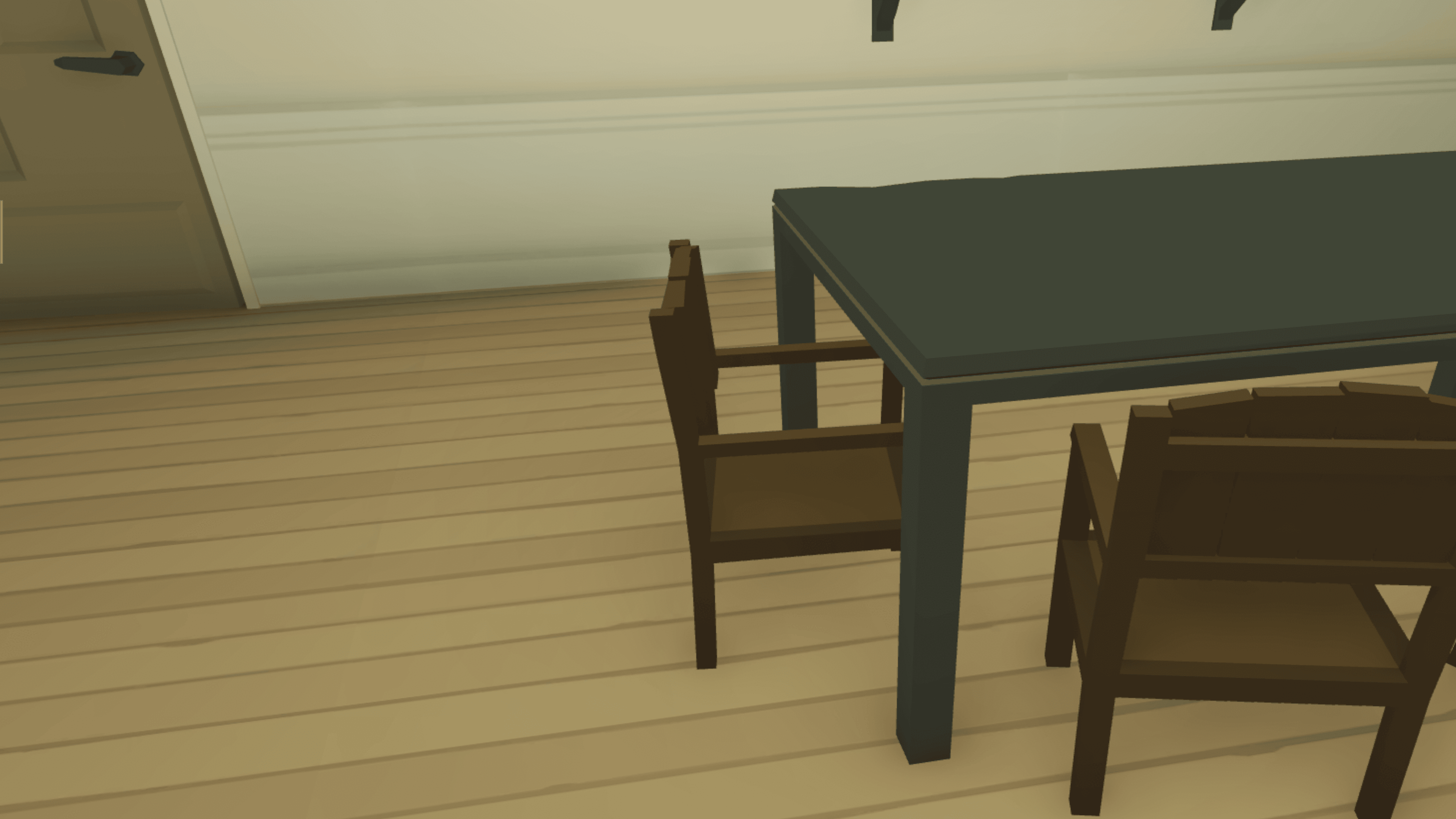}
        \caption{Easy, Light On, Left}
    \end{subfigure}
    \hfill
    \begin{subfigure}{0.32\textwidth}
        \includegraphics[width=\linewidth]{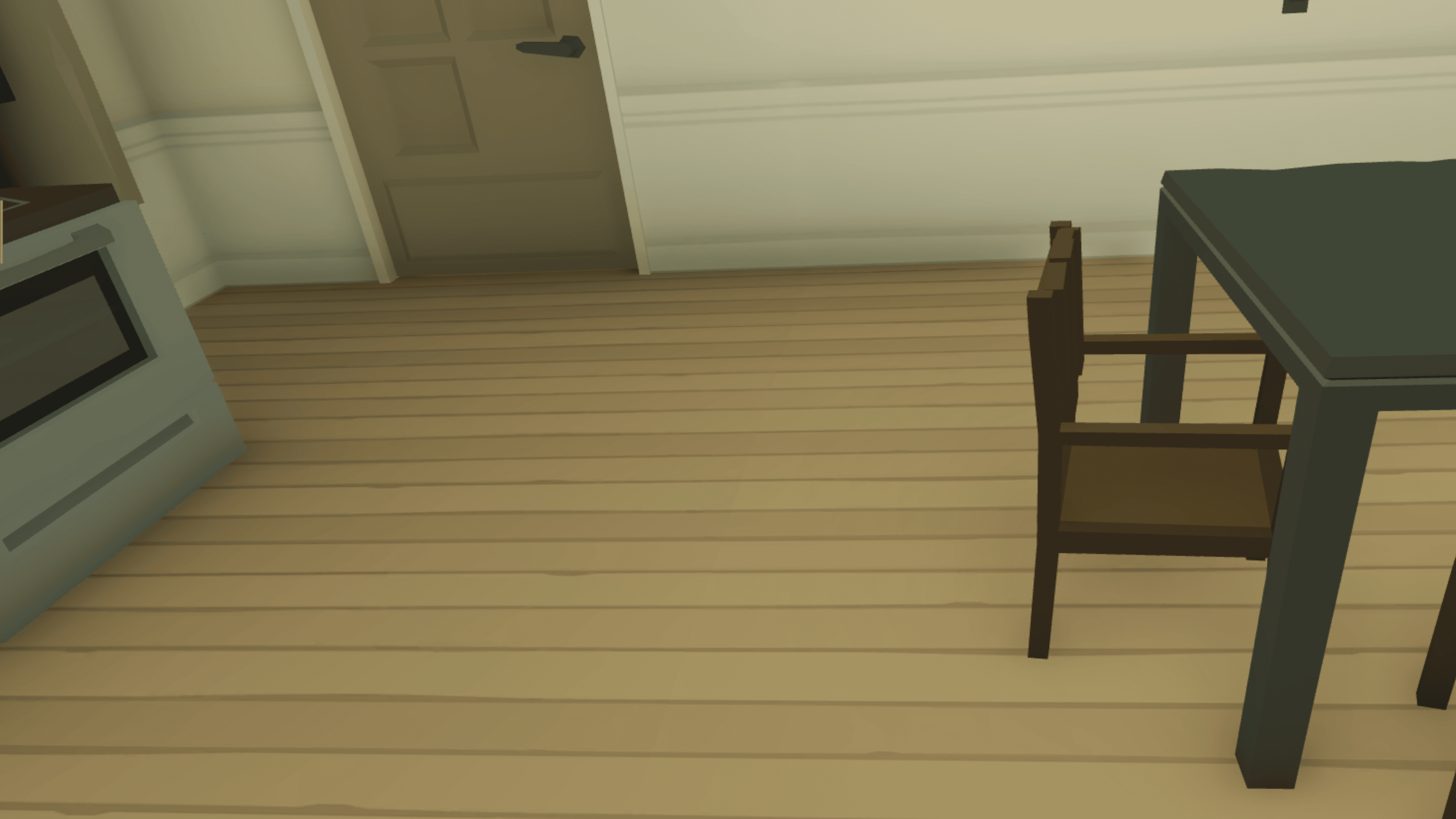}
        \caption{Easy, Light On, Far Left}
    \end{subfigure}

        \centering
    \begin{subfigure}{0.32\textwidth}
        \includegraphics[width=\linewidth]{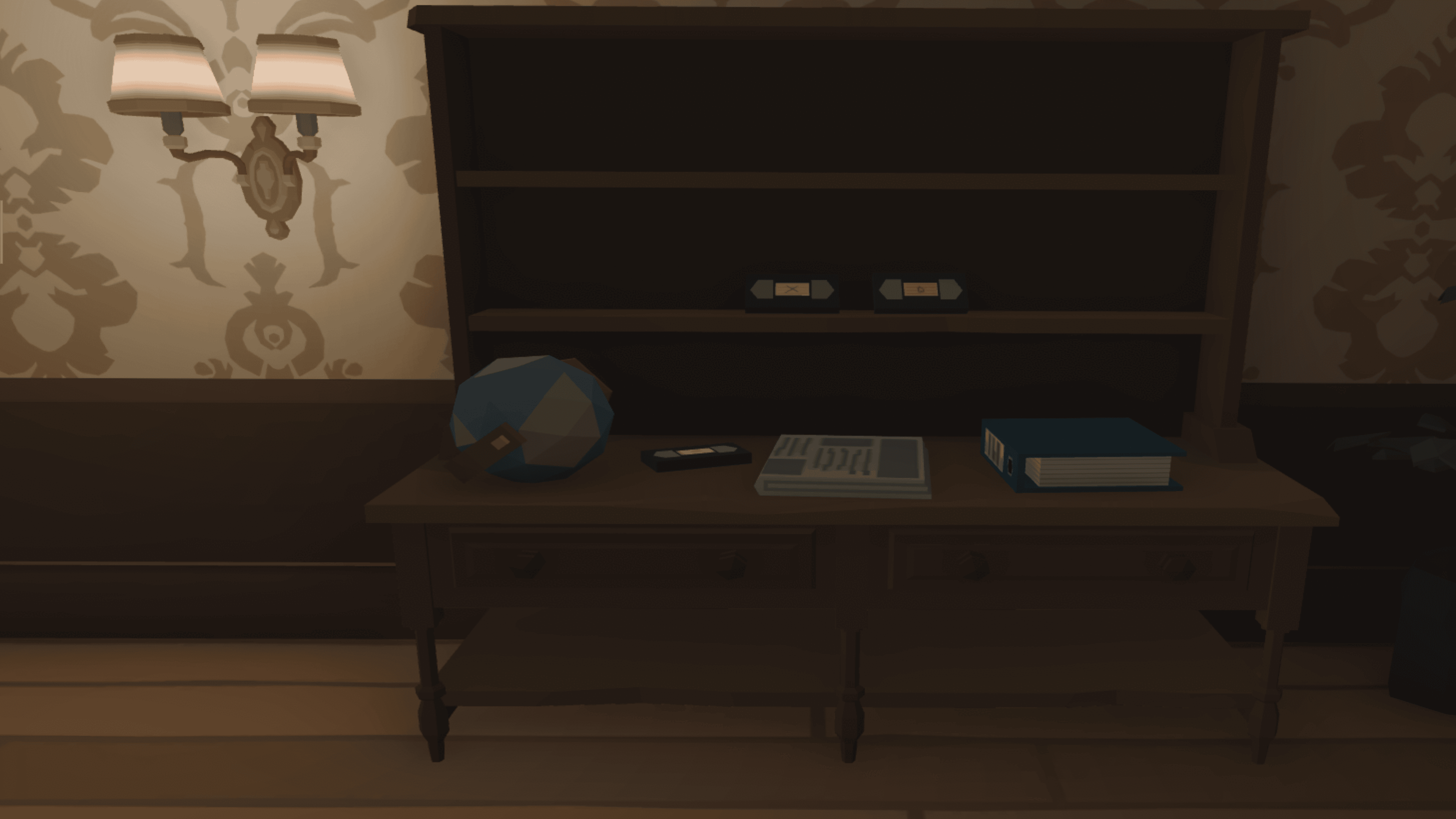}
        \caption{Med., Light On, Front}
    \end{subfigure}
    \hfill
    \begin{subfigure}{0.32\textwidth}
        \includegraphics[width=\linewidth]{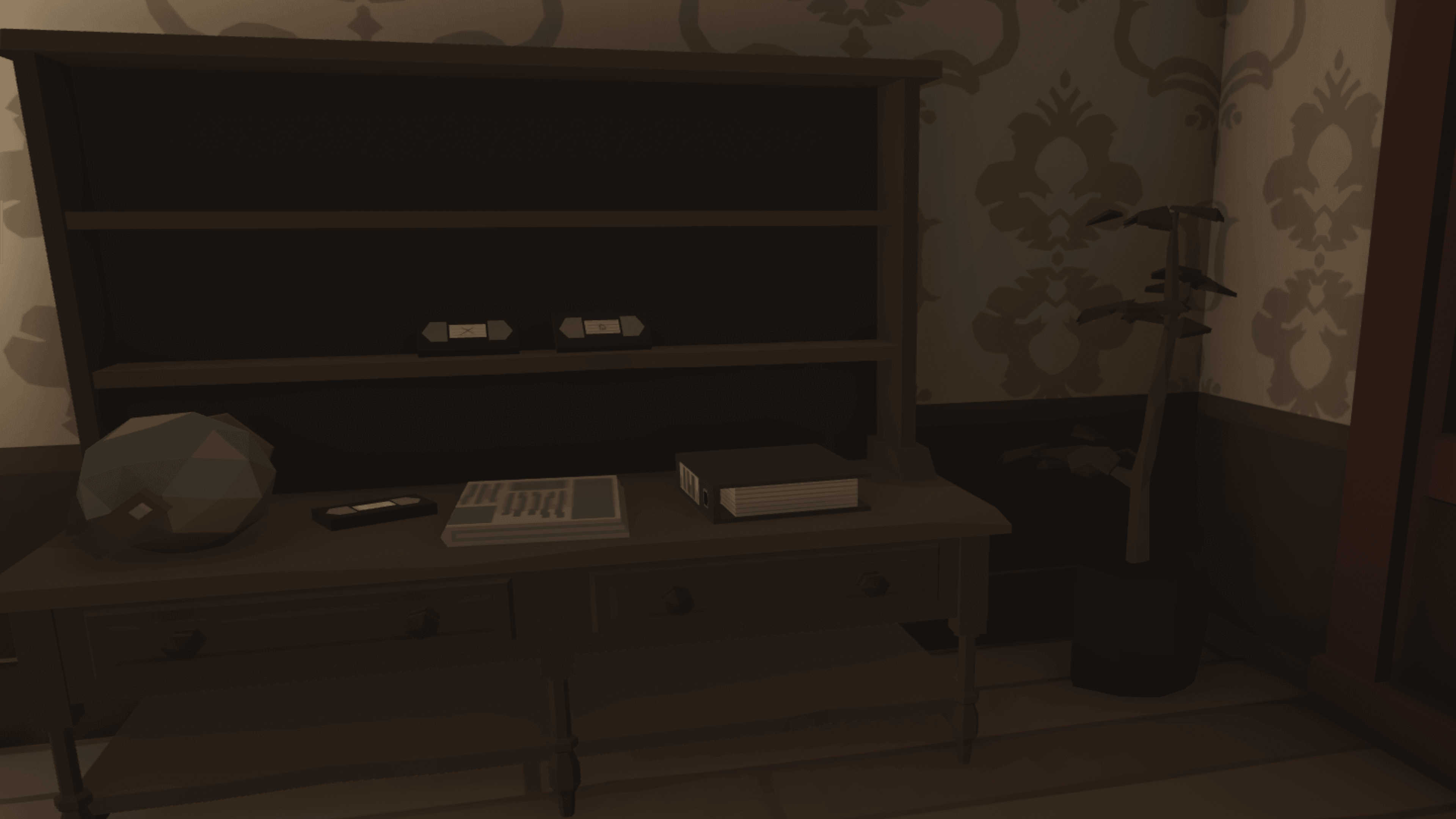}
        \caption{Med., Light On, Right}
    \end{subfigure}
    \hfill
    \begin{subfigure}{0.32\textwidth}
        \includegraphics[width=\linewidth]{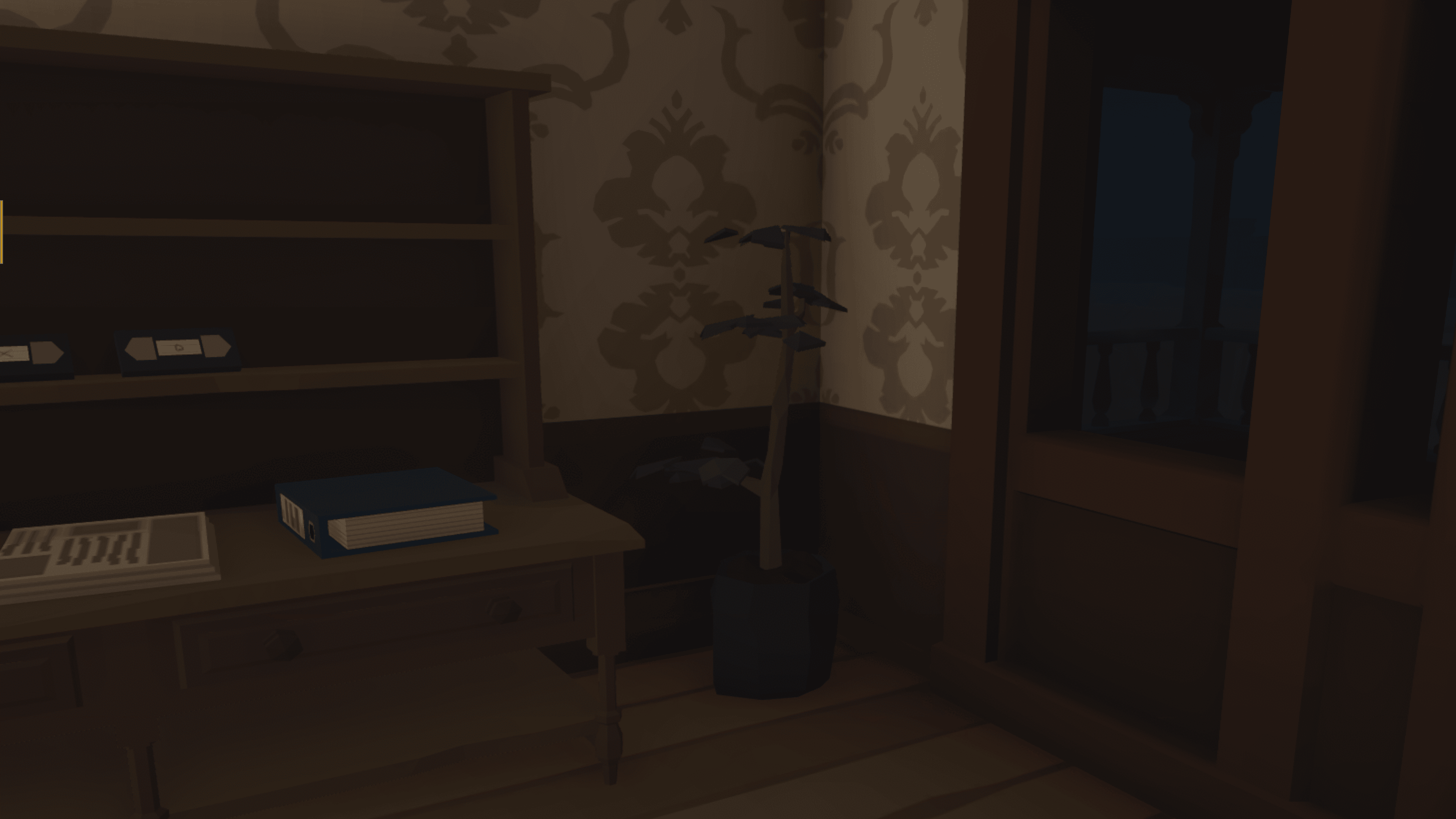}
        \caption{Med., Light On, Far Right}
    \end{subfigure}

        \centering
    \begin{subfigure}{0.32\textwidth}
        \includegraphics[width=\linewidth]{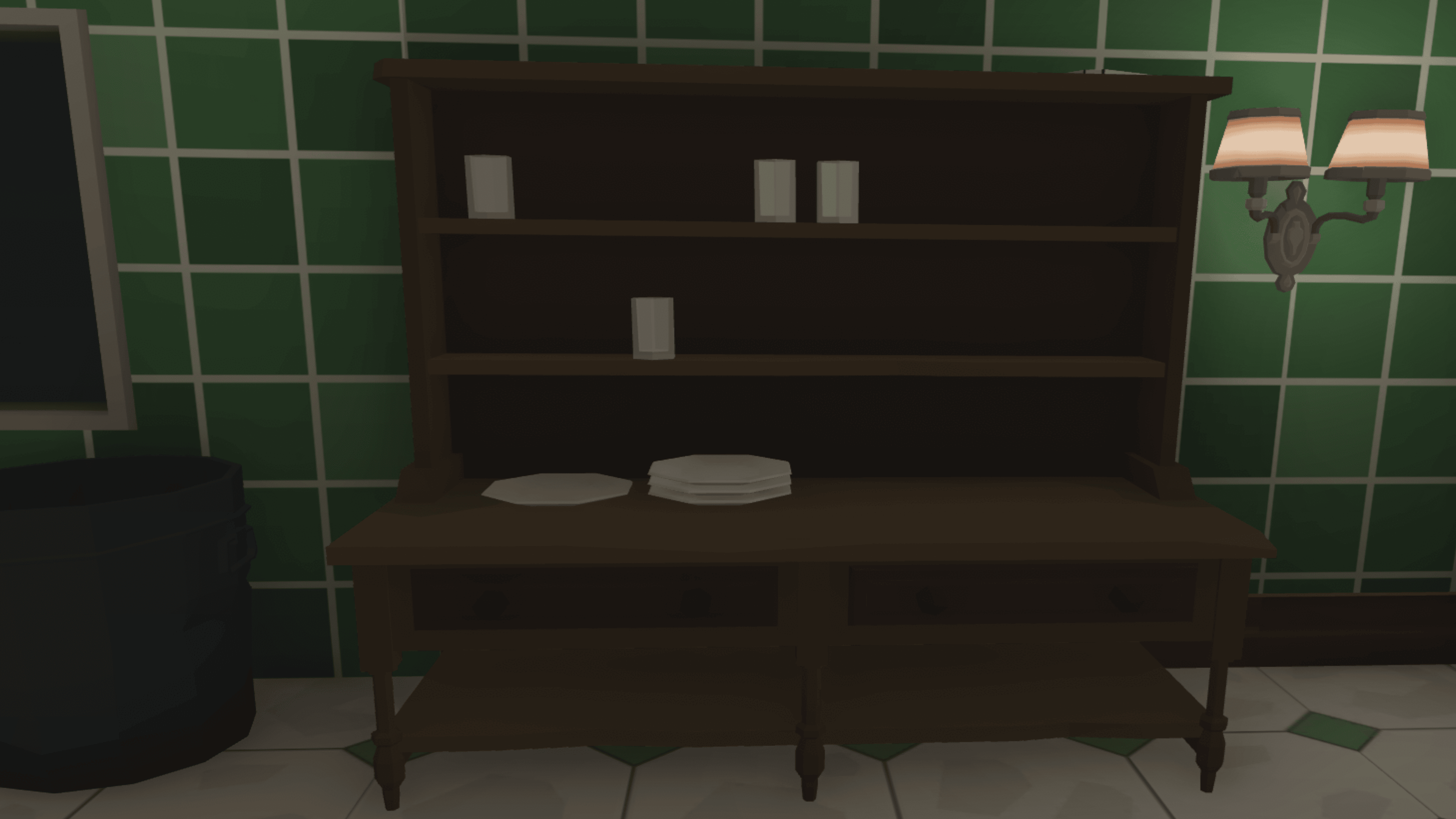}
        \caption{Hard, Light on, Front}
    \end{subfigure}
    \hfill
    \begin{subfigure}{0.32\textwidth}
        \includegraphics[width=\linewidth]{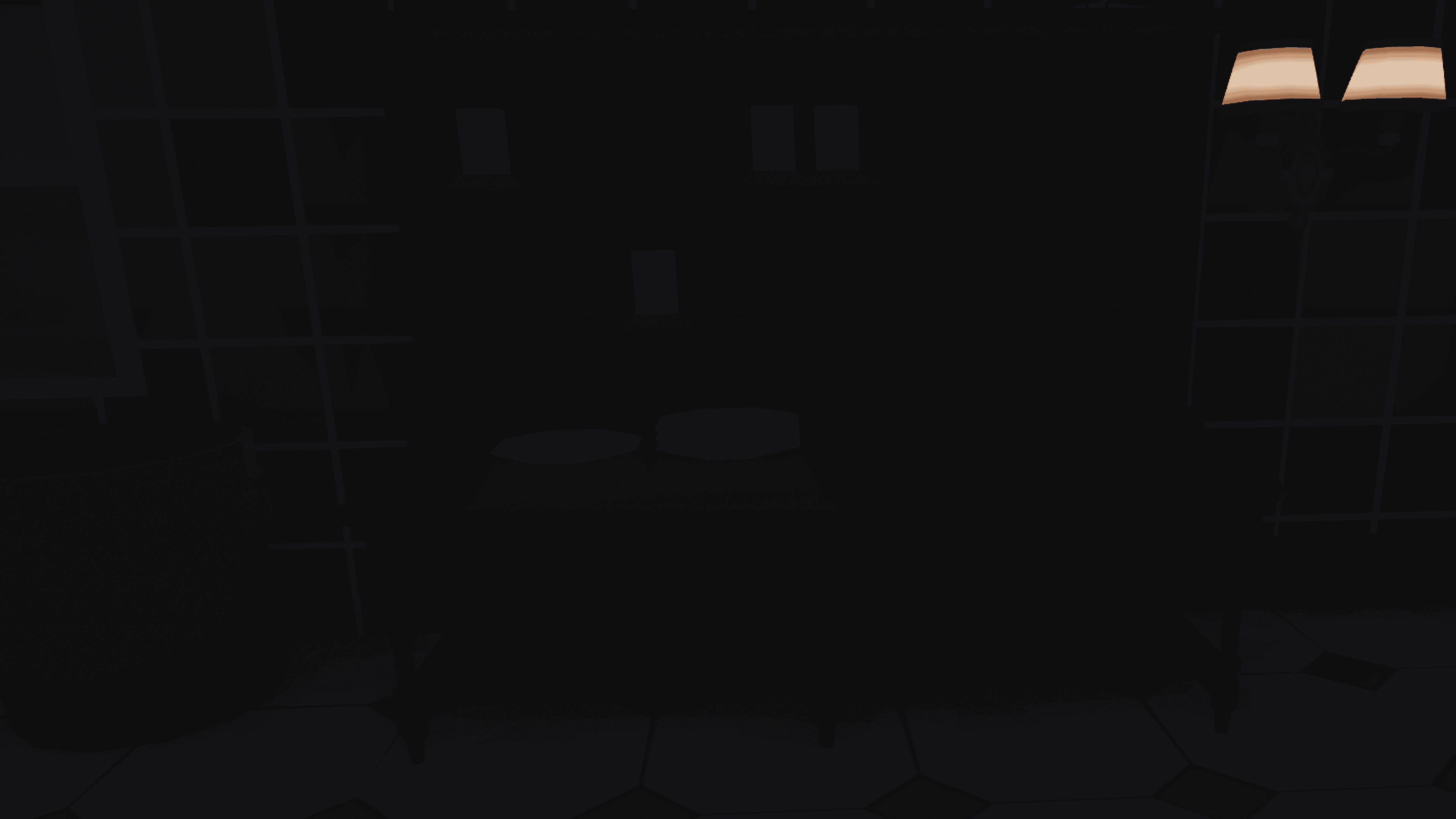}
        \caption{Hard, Light off, Front}
    \end{subfigure}
    \hfill
    \begin{subfigure}{0.32\textwidth}
        \includegraphics[width=\linewidth]{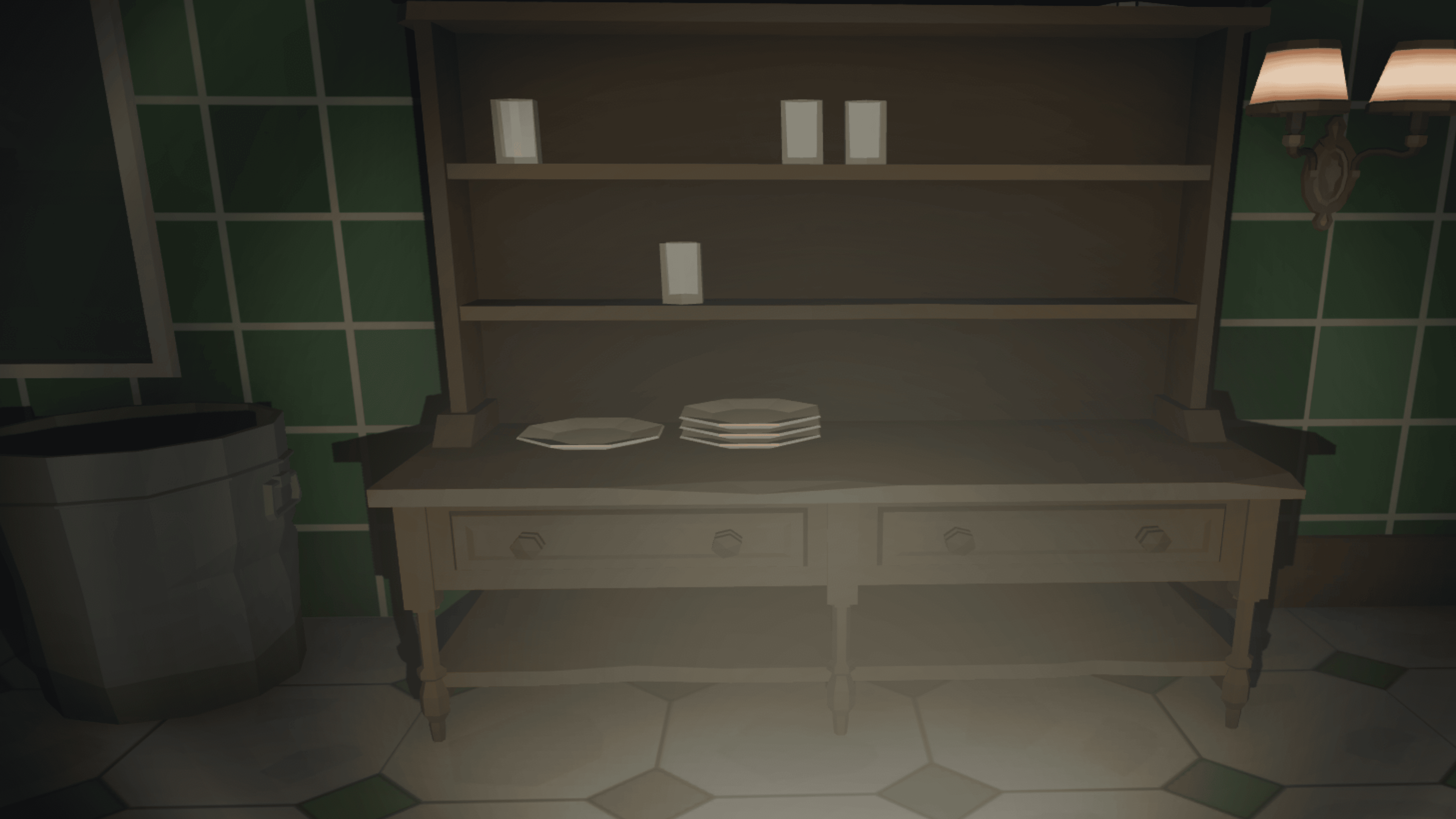}
        \caption{Hard, Flashlight, Front}
    \end{subfigure}
    \caption{Examples of Dataset Representing Diverse FOV}
    \label{fig:dataset}
\end{figure}

To support all five study experiments, we select a VR simulation game named \textit{The Break-In}, which is a thief-simulating game with a variety of virtual environments, including houses and stores. 
The game supports multiple devices and platforms, and we use Meta Quest 2~\cite{meta_quest2_getting_started} and install the game version .45 through Meta Store.
We select a house as our experimental scene and let the VR player randomly explore the rooms, such as the dining room, bathroom, and nursery.
During the exploration, we observe that different rooms contain varying numbers of entities and share distinct lighting conditions.
Therefore, to organize the Field of Views (captured screenshots), we define three levels of complexity by controlling the number of entities in them: 
\begin{itemize}[left=2em]
    \item Easy level (less than five entities),
    \item Medium level (five to ten entities),
    \item Hard level (over ten entities). 
\end{itemize}
Additionally, for each FOV, we capture its three different lighting conditions: lights on, flashlight, and light off. To further address RQ4 and RQ5, we collect four additional viewpoints~\cite{def_viewpoint} in each room by asking the VR player to turn left (30 degrees), far left (60 degrees), right (30 degrees), and far right (60 degrees).
Note that the VR screenshots usually contain two versions: \textit{left eye} and \textit{right eye}.
In this experiment, we capture the screenshots as monoscopic images.
This is because the built-in screenshot capture feature in Meta Quest 2 only takes a flat monoscopic 2D image from the default right eye’s perspective. Capturing a stereoscopic view, however, requires more effort, such as installing developer tools or third-party plugins. Moreover, different devices will require different customized solutions.
Since we want to design a more general approach that can be adopted to all kinds of VR devices, we focus more on the general approaches and thus only use the flat monoscopic 2D image.
We collect 135 screenshots (45 for each complexity level), including nine FOVs from different rooms; each FOV has three lighting conditions and five different perspectives.
Moreover, for each screenshot, we capture it with an original resolution of $3840 \times 2160$, and then compress it to $512 \times 512$, maintaining the same aspect ratio.
The total number of screenshots in our dataset is $270$.
Figure~\ref{fig:dataset} shows a few examples of our dataset representing diverse scene settings.

\subsection{Groundtruth Construction}
\label{subsec:groundtruth_construction}
To evaluate the LLM's performance and answer the research questions,
we adopt the open coding method~\cite{charmaz2006constructing} to conduct a qualitative analysis of all the results and build a labeled dataset of groundtruth based on the voting results. 
In general, Annotator 1 and Annotator 2 would first independently label each FOV, and Annotator 3 then reviews the annotations to determine whether the two labels are consistent. If the annotations were consistent, they were directly accepted as the groundtruth result. If not, Annotator 3 selects the most appropriate annotation as the groundtruth result.

\begin{figure}[hbp]
    \centering
    \begin{subfigure}{0.49\textwidth}
        \includegraphics[width=\textwidth]{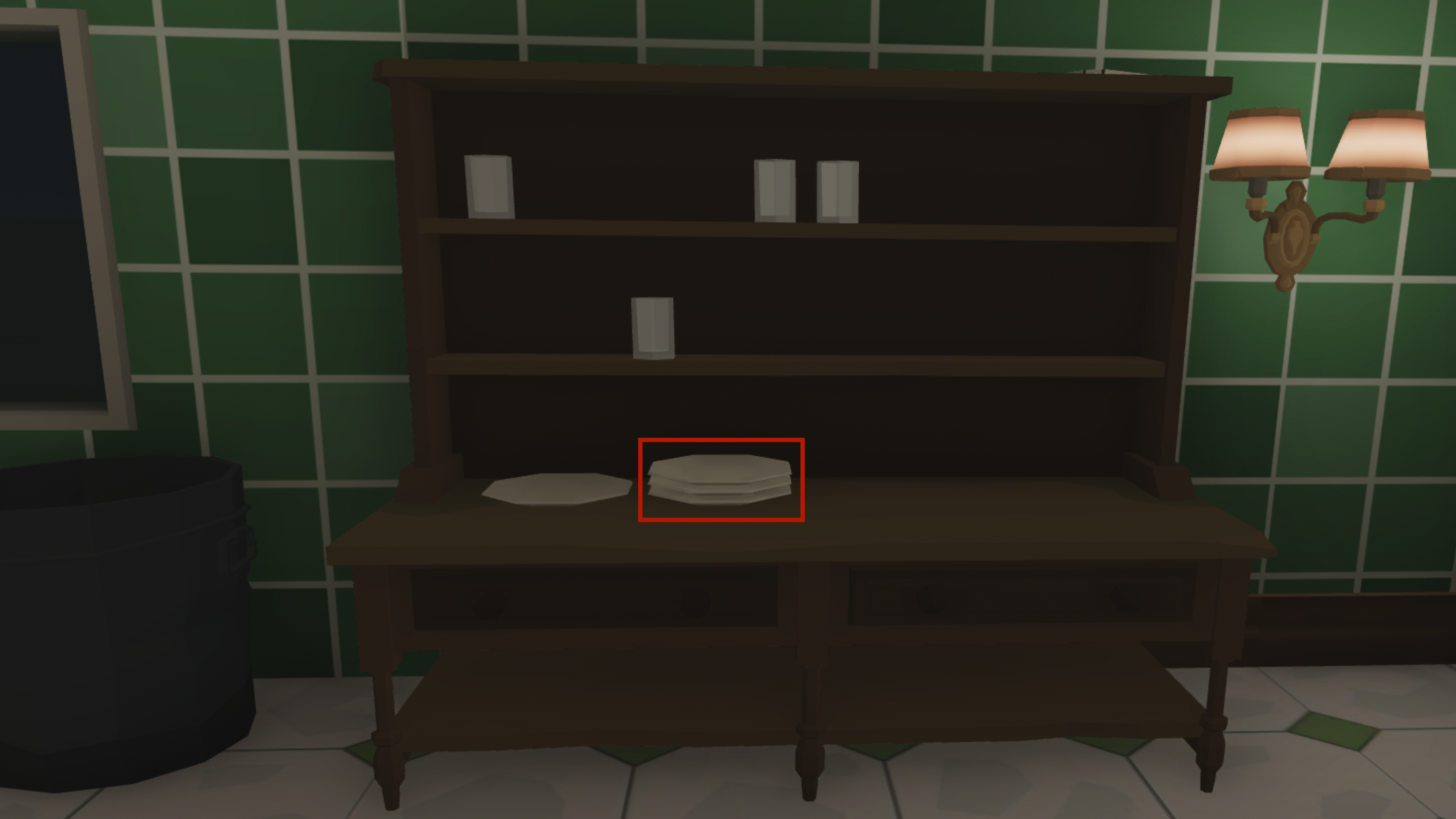}
        \caption{Example 1}
    \end{subfigure}
    \begin{subfigure}{0.49\textwidth}
        \includegraphics[width=\textwidth]{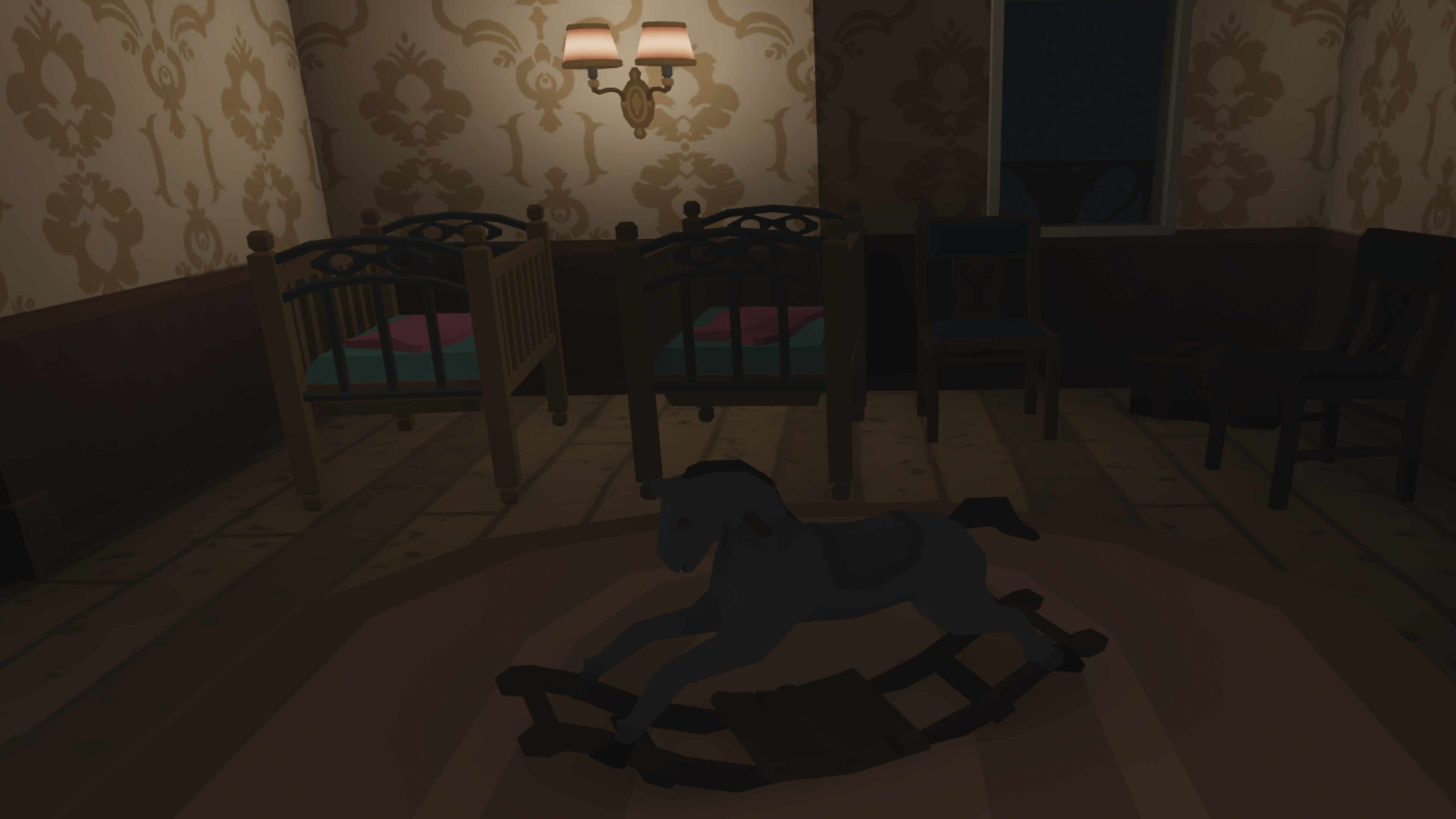}
        \caption{Example 2}
    \end{subfigure}
    \caption{Examples of Groundtruth Construction}
    \label{fig:voting_example}
\end{figure}

For instance, as depicted in Figure \ref{fig:voting_example} (a), Annotator 1 categorizes the object in the red box as \textit{plates}, while Annotator 2 identifies it as three distinct \textit{plate} objects, resulting in an inconsistent annotation. In this scenario, Annotator 3 votes the same as Annotator 2’s results because \textit{plates} encompasses multiple independent objects as a whole, disregarding the spatial relationship, which is a keypoint in the following research question. Thus, the majority voting, \textit{three plate objects}, is selected as the groundtruth.

Similarly, in Figure \ref{fig:voting_example} (b), Annotator 1 labels the scene as a \textit{nursery}, while Annotator 2 categorizes it as a \textit{children’s room}, leading to an inconsistent annotation. In such a case, Annotator 3 votes for Annotator 1’s result because the baby cribs in the image indicate that the room is for infants, making \textit{nursery} a more precise label. Accordingly, the groundtruth is the majority voting \textit{nursery}.

In RQ2, through Open Coding and Voting, we obtain a total of $216$ groundtruth labels for entities across $27$ images, representing nine different scenarios~\cite{def_scenario} under varying lighting conditions.
In RQ3, we manually label three-dimensional information for $216$ entities and create $9$ scene descriptions.
In RQ5, we manually annotate $36$ pairs of entities across different FOVs, where the entities in green and red are the same and in red and blue are different.

\section{Methodology} \label{sec:method}

\begin{figure}[ht]
  \centering
  \includegraphics[width=0.95\linewidth]{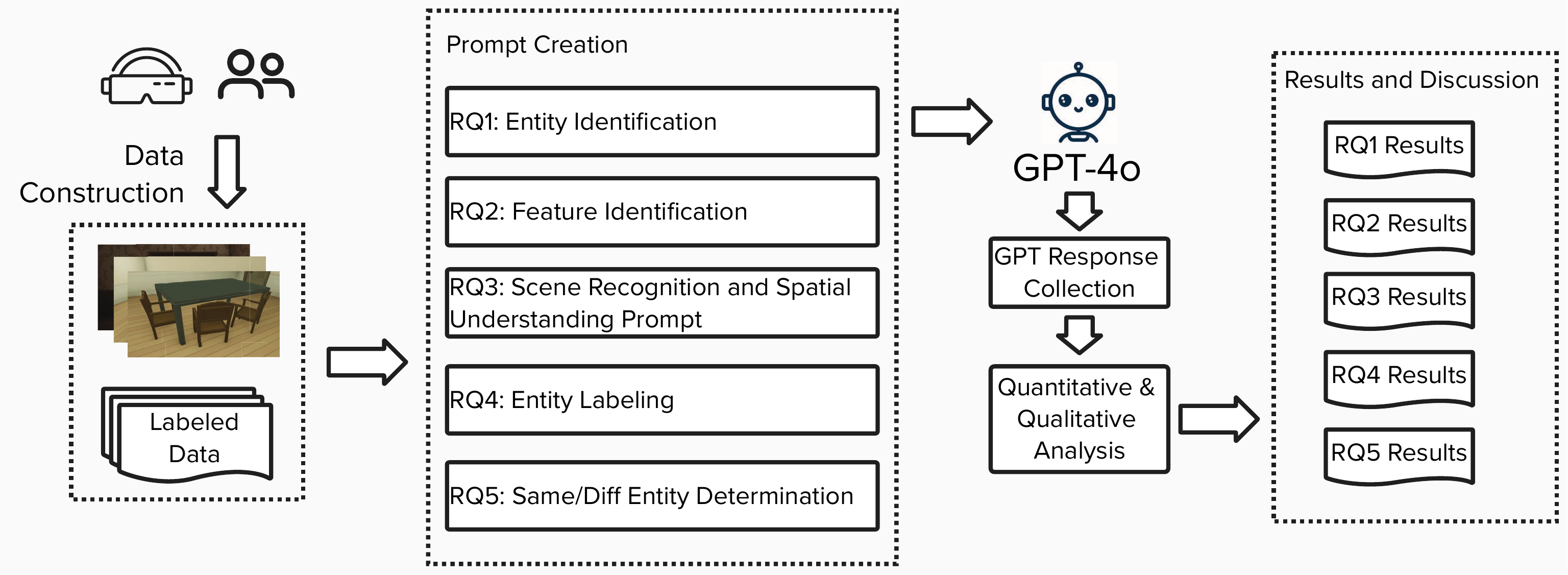}
  \caption{Overview of the Approach}
  \label{fig:overview}
\end{figure}

Figure~\ref{fig:overview} shows the overview of the study approach.
In this study, we begin by having two VR players collect raw VR screenshots. We then use the open-coding method to construct the labeled groundtruth data. Next, we create one set of prompts for each of the five research questions 
and submit them to the LLM, specifically GPT-4o, to gather responses. For each GPT-generated answer, we conduct quantitative and qualitative analysis. Finally, we present the observed performance of GPT-4o, followed by detailed discussions and potential solutions for limitations. 

The following sections describe the details of the designed experimental studies that were used to answer the research questions listed in Section~\ref{sec:intro}. 
For each study, we carefully discuss the study design and possible evaluation metrics to assess the results.

\textbf{Study Design Summary.} 
All studies were inspired by the current state-of-the-art black-box GUI exploration testing.
In particular, we focus on exploring how LLMs can help GUI exploration testing in Virtual Reality applications.
The first study (RQ1, Section~\ref{subsec:study1}) is to leverage prompt engineering to obtain more accurate answers from GPT-4o. Then, the second study (RQ2, Section~\ref{subsec:study2}) evaluates how accurately GPT-4o can describe an entity's features and what are the key features of an entity. In the third study (RQ3, Section~\ref{subsec:study3}), we aim to assess GPT-4o’s ability to understand the whole scene from a partial view (FOV) and infer the spatial relationships of entities from the user’s perspective. We then evaluate GPT-4o's capability of labeling entities on a single FOV in the fourth study (RQ4, Section~\ref{subsec:study4}).
Although entities can be successfully identified by LLMs on a single FOV, can they still be recognized through multiple FOVs? We designed the last study (RQ5, Section~\ref{subsec:study5}) to find how LLM determines the same entities through FOVs.

\subsection{Study 1: Entity Detection on single FOV} \label{subsec:study1}

\subsubsection{Experiment Design}

To address RQ1, we adopt prompt engineering to explore optimal prompts to identify entities within an arbitrary given FOV, as shown in Figure~\ref{fig:dataset}.
To evaluate different prompts' performance in general, we let the LLM repeatedly perform the same task with each prompt individually on three complexity levels of FOVs, where each level consists of a different number of entities.
We then select the best prompt based on overall performance.

\begin{table}[h!] 
    \centering
    \caption{Prompts Design for Entity Detection}
    \label{tab:study1}
    \begin{tabularx}{\linewidth}{|l|X|}
        \hline
        \textbf{Text Prompt} & \textbf{Description} \\ \hline
        Basic Prompt & What is in the picture? \\ \hline
        Designed Prompt Part 1 & Your task is to analyze the provided image and return structured data. Follow this rule strictly: List every entity individually.  \\ \hline
        Designed Prompt Part 2 & Examine the image carefully to see if there are any missing items. If there are, add them and return the complete JSON. \\ \hline
    \end{tabularx}
\end{table}

\textbf{Prompts Design.}
The multimodal input consists of a text prompt and an image input.
The image input is the FOV, a captured screenshot during the VR application exploration.
The text prompt is a short paragraph that describes the task requirements of detecting entities in the given FOVs.
We use one intuitive prompt (Basic prompt) and design one set of structured prompts (Part 1 and Part 2) with details shown in Table~\ref{tab:study1}.
The basic prompt, which is the most common type, simply provides a general query without specific instructions. Thus, we use it as the baseline.
The designed structured prompts include two parts:
Part 1 includes the detailed requirements for listing entities from the FOVs with examples to improve the entity recognition accuracy rate in FOVs.
Part 2 design is based on Chain-of-Thought (CoT) Prompting~\cite{ge2023chainthoughtprompttuning}; verification techniques are applied to further enhance the recognition rate and accuracy of entity detection in FOV. Specifically, each time, we feed the inputs and outputs from Part 1, serving as a temporary short memory. 
Ideally, the second part will guide LLM in self-correcting the mistakes that it made in the first part.
To verify if more iterations of self-corrections yield higher accuracy, we repeatedly apply Part 2 
from zero time to multiple times until the performance does not significantly increase.
We aim to find the optimal CoT prompts, including Part 1 and the suitable number of Part 2 iterations.

\subsubsection{Evaluation Metrics}
\label{sec:study1_dcem}

To determine the performance of each prompt, we compare the LLM-obtained results with groundtruth. 
We use two metrics for quantitative evaluation.
First, we use Correct Detection Rate (CDR) (Equation~\ref{eq:CDR}) to quantify the percentage of correctly detected entities relative to the total number of groundtruth entities.
Second, we use Extra Detection Rate (EDR) (Equation~\ref{eq:EDR}) to measure the proportion of falsely detected extra entities over the groundtruth, including walls, windows, and floors.
A good performance should have a higher CDR and maintain a lower EDR.

\begin{equation}
\label{eq:CDR}
CDR = \frac{Number_{CorrectDetection}}{Number_{GroundTruth}} \times 100\% 
\end{equation}

\begin{equation}
\label{eq:EDR}
EDR = \frac{Number_{ExtraDetection}}{Number_{GroundTruth}} \times 100\% 
\end{equation}

\subsection{Study 2: Identifying Core Features of an Entity} \label{subsec:study2}

\subsubsection{Experiment Design}
To answer RQ2, we first explore the state-of-the-art in different disciplines to understand what features are generally used to represent an entity. And then, we contextualize these features to make them fit our experiment setup.
Lastly, we adopt the optimized prompts from RQ1 for entity identification and add prompts to have the LLM describe contextualized features.  

\textbf{Feature Collection from Literature.}
We collect feature candidates in two aspects: first is the current AI model's practice of identifying features from image processing.
The second one is how human beings recognize entities cognitively.
For the first aspect, we review $17$ papers, with the majority from Computer Vision, and identify features of color~\cite{diplaros2003color, van2009evaluating, gevers2000pictoseek, giesel2010color, duong2010shape, ge2022contributions, kasaei2021investigating, olkkonen2008color, adelson2001seeing, cremers2007review, jumi2021performance}, shape~\cite{diplaros2003color, van2009evaluating, gevers2000pictoseek, hertzmann2003shape, duong2010shape, ge2022contributions, kasaei2021investigating, kadir2015new, jumi2021performance, ferrari2010images, olkkonen2008color, cremers2007review, leone2007shadow}, texture and material~\cite{kadir2015new, ge2022contributions,jumi2021performance, adelson2001seeing, cremers2007review, leone2007shadow, giesel2010color, hertzmann2003shape, duong2010shape, sharan2009perception, armi2019texture}, placement (location and orientation)~\cite{diplaros2003color}, and lighting~\cite{hertzmann2003shape, olkkonen2008color}.
For the second aspect, we assume that the end user’s understanding of the VR application directly relates to FOVs. To explore this, we focus on theories of how humans perceive and understand entities in the real world. Specifically, we review key hypotheses from cognitive science~\cite{Leow1997VisualSchemas, TREISMAN198097} and identify the relevant features of color, orientation, spatial frequency, brightness, and visual schemas (both component and context schemas).

\textbf{Feature Contextualization.}
In total, we collect nine cumulative features: \textit{color, shape, texture, material, brightness, orientation, spatial frequency, component schemas, and context schemas}. 
We combine orientation and location to form a compound feature, which we refer to as placement. 
And we merge context schemas with placement.
We further merge the component schemas with shape, and merge texture with material since they refer to similar concepts.
We exclude spatial frequency because the feature identification results require manual verification, and spatial frequency is not interpretable by human testers.
The final five features we selected are \textbf{\textit{color}}, \textbf{\textit{placement} (location and orientation)}, \textbf{\textit{lighting}}, \textbf{\textit{texture and materials}}, and \textbf{\textit{shape}}.

\begin{figure}
  \centering
  \includegraphics[width=0.9\linewidth]{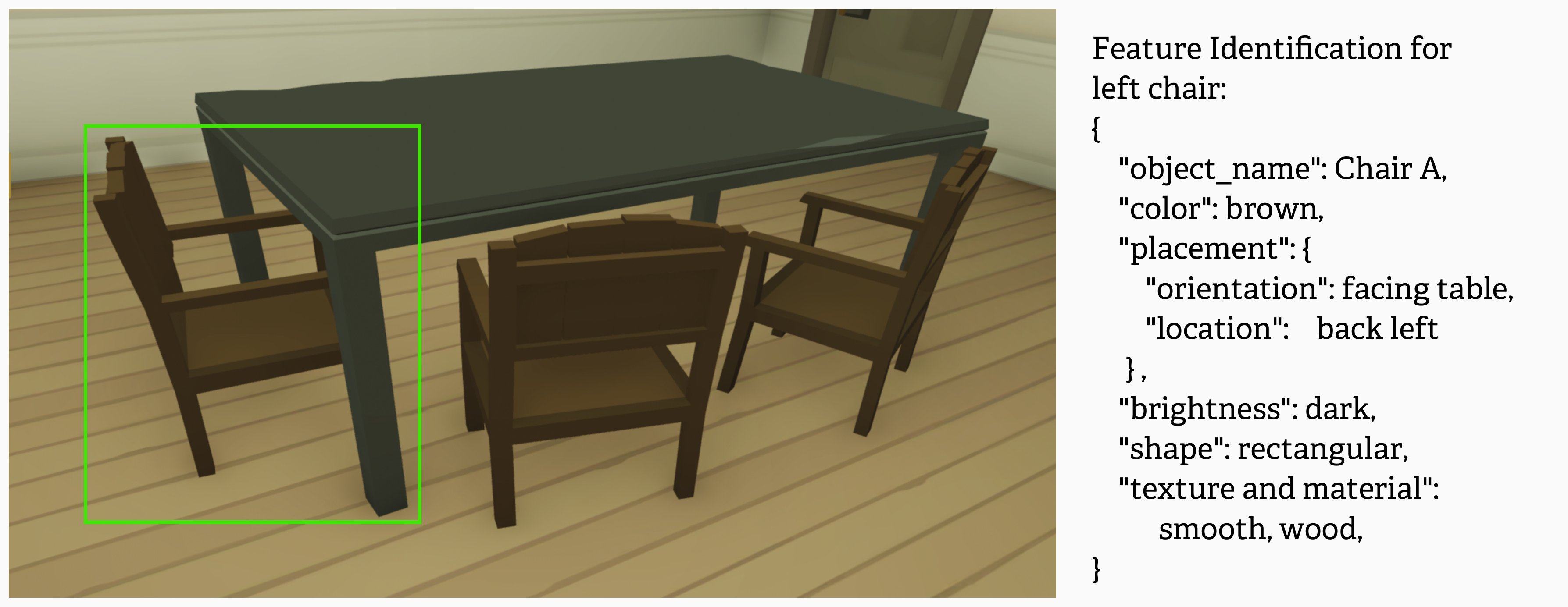}
  \caption{Example of Feature Identification}
  \label{fig:study2_feature_identification_new}
\end{figure}

\textbf{Feature Identification.}
First, we use the optimized prompts from RQ1 to detect the entities on the given FOVs.
For the detected entity, we further ask the LLM to describe the value of the five features (prompt shown in Table~\ref{tab:study2}).
For example, Figure~\ref{fig:study2_feature_identification_new} shows an example of a FOV and the identified features of an entity inside. The detected entity is the chair on the left, and the identified features with their values are listed on the right side.
Note that there might be duplicated entities on the same FOV, which is normal for VR applications due to reusing assets. In this case, we ask the model to name different entities and append the names with different letter identifiers.
In this FOV, we collect three different feature identification results for entity ``Chair A'', entity ``Chair B'', and entity ``Chair C'', respectively.

\begin{table}[h!] 
    \centering
    \caption{Prompts Design for Feature Identification (New prompts marked \textbf{bold}.)}
    \label{tab:study2}
    \begin{tabularx}{\linewidth}{|l|X|}
        \hline
        \textbf{Text Prompt} & \textbf{Description} \\ \hline
        Designed Prompt Part 1 & Your task is to analyze the provided image and return structured data. Follow this rule strictly:
        
        \begin{enumerate}
            \item List every object individually.
            \item \textbf{For each object, provide:}
            \begin{itemize}
                \item \textbf{\textcolor{blue!60!black}{Name}: The object name with a unique identifier (e.g., 'Chair A', 'Chair B').}
                \item \textbf{\textcolor{blue!60!black}{Color}:} 
                \item \textbf{\textcolor{blue!60!black}{Placement (Orientation and location)}:} 
                \item \textbf{\textcolor{blue!60!black}{Lighting}:} 
                \item \textbf{\textcolor{blue!60!black}{Shape}:} 
                \item \textbf{\textcolor{blue!60!black}{Texture and Material}:} 
            \end{itemize}
        \end{enumerate}\\ \hline
        Designed Prompt Part 2 & Examine the image carefully to see if there are any missing items. If there are, add them and return the complete JSON. \\ \hline
    \end{tabularx}
\end{table}

\subsubsection{Evaluation Metrics}
As discussed in Section \ref{subsec:groundtruth_construction}, we adopt the open coding method to first manually label the observed entities and their features for each FOV.
Note that all three annotators do not see the LLM answers before listing all the groundtruth values.
Then, we follow a major voting strategy between the LLM results and the groundtruth values to decide whether a match exists. Ultimately, we calculate the Feature Identification Rate (FIR) for both overall and for individual features using Equation~\ref{eq:fir}.

\begin{equation} \label{eq:fir}
FIR = \frac{Number_{MatchedFeatures}}{Number_{TotalFeatures}} \times 100\% 
\end{equation}

\subsection{Study 3: Scene and spatial analysis on single FOV} \label{subsec:study3}

\subsubsection{Experiment Design}
In the VR world, the user usually locates an entity by understanding the virtual environment first and then referring to its spatial relationship for navigation.
For example, to locate the left chair in Figure~\ref{fig:study2_feature_identification_new}, a user will first recognize the environment as a dining room, and then the target chair is on the left-hand side.
Given the disparity between the scenes, the user is unlikely to identify this chair as another chair from the living room.
To address RQ3, we measure the performance of LLM on recognizing the scene and understanding the spatial relationships between a target entity and the user through FOV analysis.

\textbf{Scene Recognition} helps future testing tasks to understand if a user is still in the same scene even when the FOV changes. It also provides the scene context for all the identified entities.
Moreover, visual schemas~\cite{Leow1997VisualSchemas} discuss the importance of context schemas, for example, an image with a car and sidewalk probably shows a street view instead of a garage.
In this study, we want to measure how well the LLM can recognize scenes from different FOVs, especially when comparing FOVs taken from different perspectives within the same room.

\begin{figure}[hbt!]
  \centering
  \includegraphics[width=0.9\linewidth]{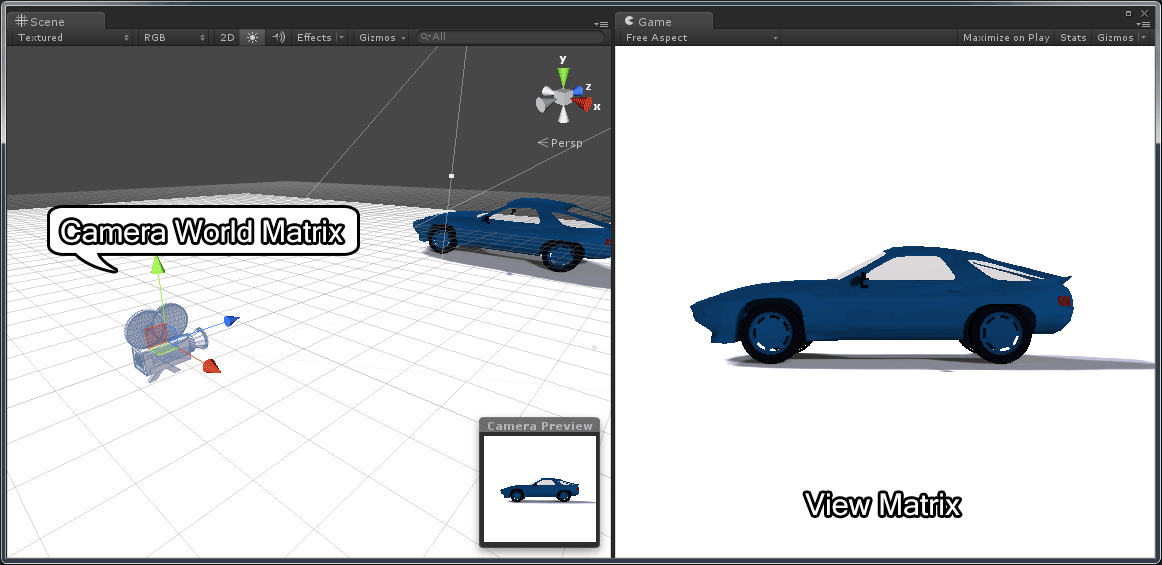}
  \caption{Camera World Matrix \textit{vs.} View Matrix}
  \label{fig:study3_spatial}
\end{figure}

\begin{table}[!ht] 
    \centering
    \caption{Prompts Design for Scene and Spatial Analysis (New prompts marked \textbf{bold}.)}
    \label{tab:study3}
    \begin{tabularx}{\linewidth}{|p{3.5cm}|X|}
        \hline
        \textbf{Text Prompt} & \textbf{Description} \\ \hline
        Designed Prompt Part 1 + Unstructured Output & Your task is to analyze the provided image and return structured data. Follow these rules strictly:

        \begin{enumerate}
            \item \textbf{Scene Type:}
            \begin{itemize}
                \item \textbf{Provide the type of scene (e.g., dining room, office).}
                \item List every entity individually.
            \end{itemize}
            \item \textbf{For each entity, provide:}
            \begin{itemize}
                \item Name: The entity name with a unique identifier (e.g., 'Chair A', 'Chair B').
                \item \textbf{Spatial Relationship:}
            \end{itemize}
        \end{enumerate}
        \\ \hline
        Designed Prompt Part 1  + Structured Output & Your task is to analyze the provided image and return structured data. Follow this rule strictly:
        
        \begin{enumerate}
            \item \textbf{Scene Type:}
            \begin{itemize}
                \item Provide the type of scene (e.g., dining room, office).
                \item List every entity individually.
            \end{itemize}
            \item \textbf{For each entity, provide:}
            \begin{itemize}
                \item \textbf{Name:} The entity name with a unique identifier (e.g., 'Chair A', 'Chair B'). 
                \item \textbf{Spatial Relationship:}
                \begin{itemize}
                    \item \textbf{\textcolor{blue!60!black}{Horizontal}:} Specify left, middle, or right.
                    \item \textbf{\textcolor{blue!60!black}{Vertical}:} Specify up, down, or center.
                    \item \textbf{\textcolor{blue!60!black}{Depth}:} Specify near, mid-range, or far.
                \end{itemize}
            \end{itemize}
        \end{enumerate}\\ \hline
        Designed Prompt Part 2 & Examine the image carefully to see if there are any missing items. If there are, add them and return the complete JSON. \\ \hline
    \end{tabularx}
\end{table}

\textbf{Spatial Understanding.}
To systematically suggest navigation information, we reuse the best-performed prompts from RQ1 and add new prompts to ask LLM to \textit{structurally} report its understanding of the spatial relationship between the user of the FOV and the target entities (prompt shown in Table~\ref{tab:study3}).
Note that there are an unlimited number of ways to organize entities in a space. 
We ask the model to spatially organize the entities in three dimensions: horizontal, vertical, and depth.
We pick these three dimensions because they are closely related to the 3D space X (horizontal), Y (vertical), and Z (depth) axes. VR applications often use these 3D positions~\cite{UnityPositionConstraint} to represent the spatial relationship between the user and the entity.
As shown in Figure~\ref{fig:study3_spatial}, which is from a 3D game programming tutorial~\cite{3dgep2023viewmatrix}, the developer uses a view matrix to transform a model’s vertices from world space to view space.
On the other hand, correctly understanding the spatial locations in the camera view~\cite{def_camera_view} will also help infer the actual 3D locations (World Matrix) during the testing. 
To better regularize the LLM results, we define the possible output values for the three directions: 
1) \textit{left}, \textit{right}, and \textit{middle} for horizontal; 
2) \textit{up}, \textit{down}, and \textit{center} for vertical; 
3) \textit{far}, \textit{near}, and \textit{mid-range} for depth.

\subsubsection{Evaluation Metrics}
Similar to Study 2 in Section~\ref{subsec:study2}, we utilize the open coding method to
evaluate the LLM-obtained results of scene recognition, because there are no dedicated answers to describe a virtual environment. Equation~\ref{eq:scr} shows how the Scene Recognition Rate (SRR) is calculated.

\begin{equation} \label{eq:scr}
SRR = \frac{Number_{CorrectRecog.}}{Number_{TotalScenes}}\times 100\% 
\end{equation}

For spatial understanding, each LLM output provides three answers corresponding to each dimension.
Each dimension is given equal weight to ensure fairness in its consideration.
We define the Spatial Understanding Rate (SUR) in Equation~\ref{eq:srr} to measure the overall correctness, where $ N $ denotes the total number of entities. $ R_{H_i} $, $ R_{V_i} $, and $ R_{D_i} $ represent the matching scores for horizontal, vertical, and depth relationships in scene $ i $, respectively. These scores are binary, which means $R_{H_i}, R_{V_i}, R_{D_i} \in \{0, 1\}$, $ 1 $ indicates a match, and $ 0 $ indicates a mismatch. The term $ \frac{R_{H_i} + R_{V_i} + R_{D_i}}{3} $ calculates the average match rate for the three spatial relationships for entity $ i $.

\begin{equation} 
\label{eq:srr}
SUR = \frac{1}{N} \sum_{i=1}^N \left( \frac{R_{H_i} + R_{V_i} + R_{D_i}}{3} \right) \times 100\%
\end{equation}

\subsection{Study 4: Entity Labeling on single FOV} 
\label{subsec:study4}

\subsubsection{Experiment Design}
In RQ4, we aim to investigate the ability of LLM to virtually label identified entities from a given FOV and to determine if this labeling aligns with the feature recognition results from RQ2.
For example, if the LLM identifies a blue mug on the countertop, can it label the mug virtually by circling the target mug on the FOV screenshot?
Or, can LLM accurately provide the visual information, such as coordinates, that could assist in
entity detection on images?
In this experiment, we revisit all the FOVs with the detected entities and their descriptions. 
Then, we design prompts (detailed in Table~\ref{tab:study4}) for the LLM to 1) draw bounding boxes around these entities and 2) output the bounding box coordinates.
Eventually, we qualitatively check if the target entities have been successfully labeled.

\begin{table}[!h] 
    \centering
    \caption{Prompts Design for Entity Labeling}
    \label{tab:study4}
    \begin{tabularx}{\linewidth}{|p{3.8cm}|X|}
        \hline
        \textbf{Text Prompt} & \textbf{Description} \\ \hline
        Prompt for Approach 1 \& 2 & Your task is to analyze the provided image and return the image with each identified entity enclosed in a bounding box.
        \\ \hline
        Prompt for Approach 3 \& 4 & Your task is to analyze the provided image and return structured data. Follow these rules strictly:

        \begin{enumerate}
            \item \textbf{Scene:}
            \begin{itemize}
                \item List every entity individually.
            \end{itemize}
        
            \item \textbf{For each entity, provide:}
            \begin{itemize}
                \item \textbf{Name:} The entity name with a unique identifier (e.g., 'Chair A', 'Chair B').
                \item \textbf{Coordinates:}
                \begin{itemize}
                    \item \textbf{left\_top\_coordinate\_x}
                    \item \textbf{left\_top\_coordinate\_y}
                    \item \textbf{right\_bottom\_coordinate\_x}
                    \item \textbf{right\_bottom\_coordinate\_y}
                \end{itemize}
            \end{itemize}
        \end{enumerate}

        Examine the image carefully to see if there are any missing items. If there are, add them and return the complete JSON. 
        
        \\ \hline
    \end{tabularx}
\end{table}

\begin{figure}[h]
    \centering
    \hfill
    \begin{subfigure}[b]{0.7\textwidth}
        \includegraphics[width=\linewidth]{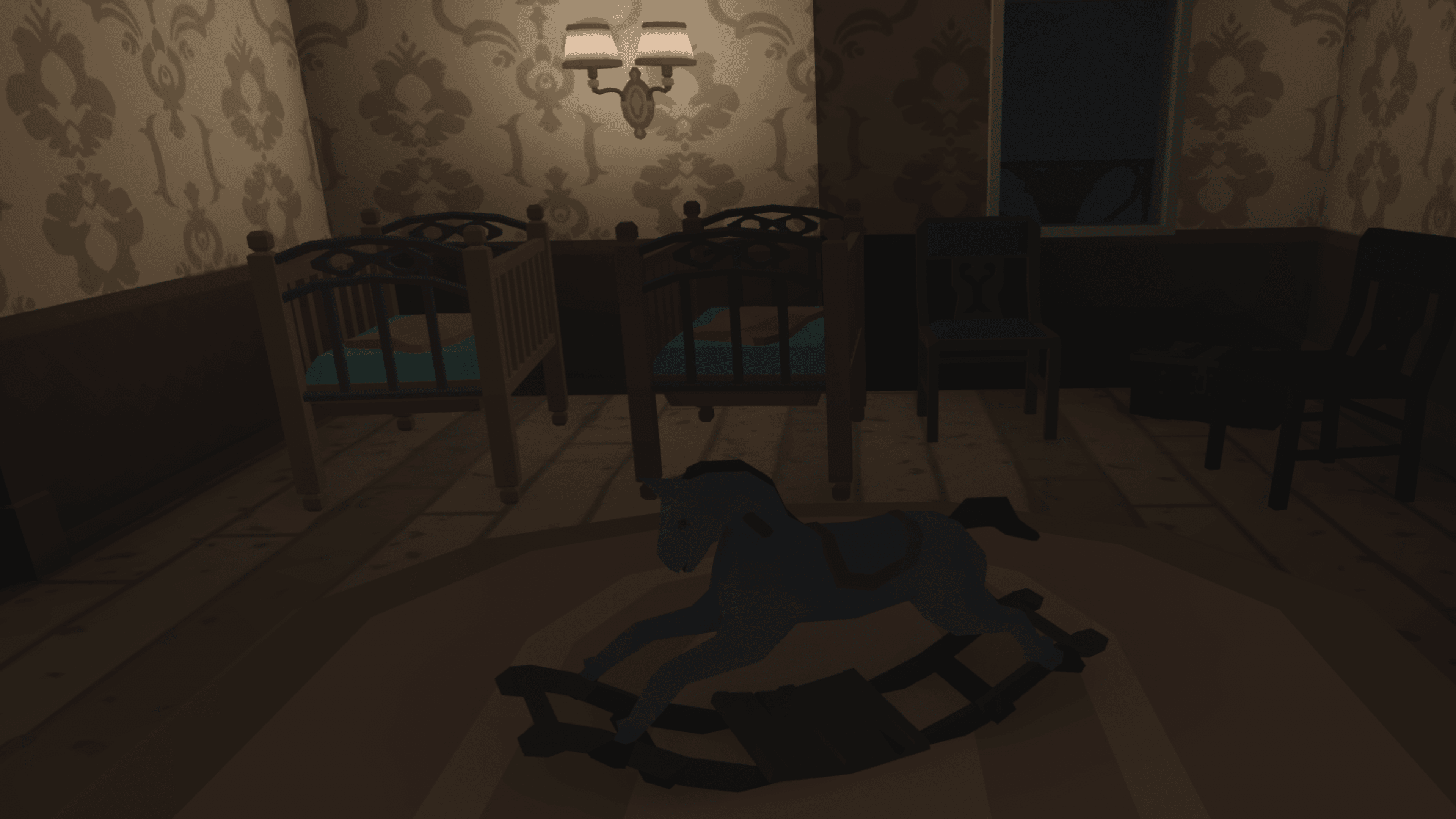}
    \end{subfigure}
    \hfill
    \begin{subfigure}[b]{0.28\textwidth}
        \includegraphics[width=\linewidth]{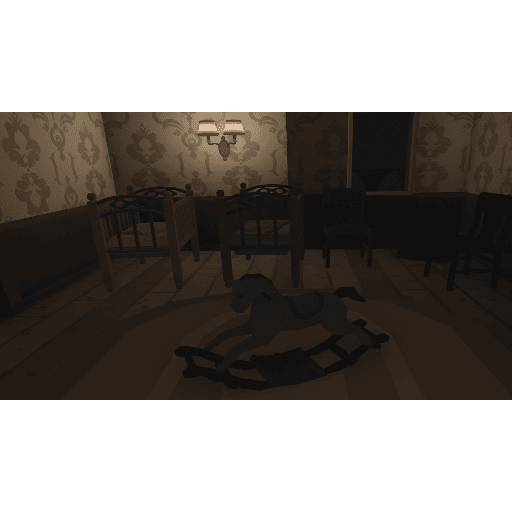}
    \end{subfigure}
    
    \caption{Example of Raw Image (Left) and User-Resized Image (Right)}
    \label{fig:original_resized_image}
\end{figure}

\textbf{Proposed Approaches.}
GPT-4o is a multimodal LLM that can handle diverse types of inputs, including text and images.
However, when exploring potential approaches for processing image input, we recognize that auto-resizing could impact our experiments. Specifically, images greater than $512 \times 512$ will be preprocessed (compressed or cropped into tiles) by GPT-4o\cite{openai_vision_limitations}. 
Consequently, we plan four approaches to address the influence of auto-resizing, aiming to mitigate the influence caused by the technique implementations. Figure~\ref{fig:original_resized_image} shows an example of a raw screenshot with an original resolution of $3840 \times 2160$ and a user-resized screenshot with a reduced resolution of $512 \times 512$ while maintaining the same aspect ratio.

\begin{itemize}
    \item Approach 1: Directly upload the raw screenshots to the LLM, and request the return of labeled images with entity annotations.
    \item Approach 2: Upload the user-resized $512 \times 512$ screenshots to the LLM, and request the return of labeled images with entity annotations.
    \item Approach 3: Upload the raw screenshots to the LLM, request the return of coordinates of the annotated entities, and use the coordinates to generate the final labeled images locally with OpenCV.
    \item Approach 4: Upload the user-resized $512 \times 512$ screenshots to the LLM, request the return of coordinates of the annotated entities, and use the coordinates to generate the final labeled images locally with OpenCV.
\end{itemize}

\subsubsection{Evaluation Metrics}

To evaluate the labeling results, we qualitatively observe the labels and the target entities on each FOV.
In particular, we project two possible scenarios for the labeling results:
1) Scenario 1: the bounding box encompasses the entire target entity;
2) Scenario 2: the bounding box shifts away from the target entity. 
We are especially interested in the number of Scenario 1 labelings.

\subsection{Study 5: Entity Determination on multiple FOVs} \label{subsec:study5}

\subsubsection{Experiment Design}

In VR apps, any motion (i.e., walking, jumping, or turning) made by users will cause the FOVs to update immediately. 
In RQ5, we investigate how LLM can help determine whether two entities are the same target during dynamic exploration with changing FOVs.
As discussed in Section~\ref{subsec:fov_creation}, we ask VR players to turn around from the original views without moving and capture FOVs in four other positions: left position, far left position, right position, and far-right position to create FOV pairs. 
We send all these FOV pairs with different view angles and ask LLM to determine whether seemingly identical entities are the same instance of the target across FOVs. 
Moreover, as FOV changes, the entity's visual representation, such as shape and placement, also changes. To explore this question in-depth, we conduct a controlled study and measure how different core features (identified in Section~\ref{subsec:study2}) will impact the 
performance of entity determination on multiple FOVs.

\begin{figure}
    \centering
    \begin{subfigure}{0.48\textwidth}
        \includegraphics[width=\linewidth]{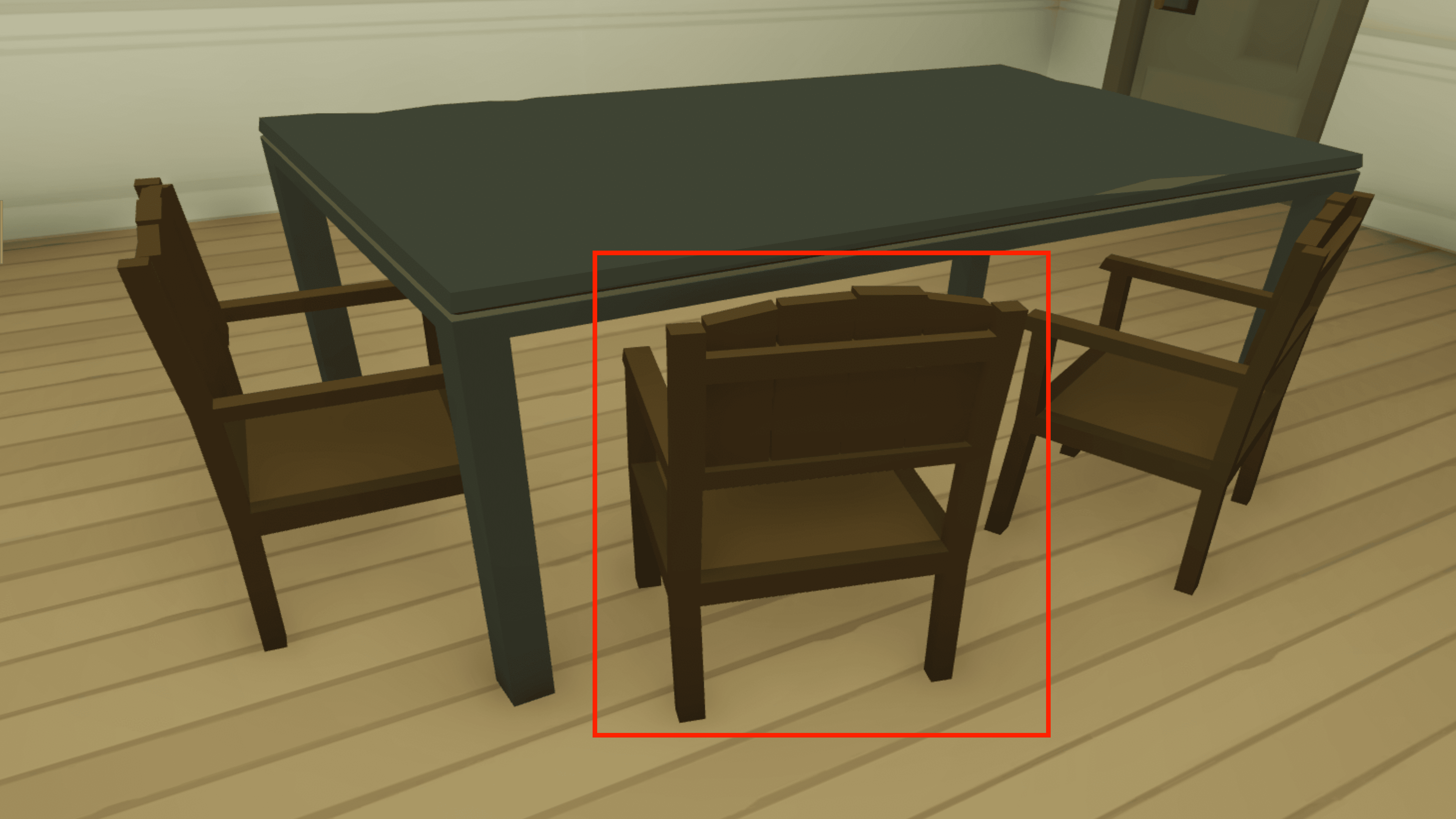}
    \end{subfigure}
    \begin{subfigure}{0.48\textwidth}
        \includegraphics[width=\linewidth]{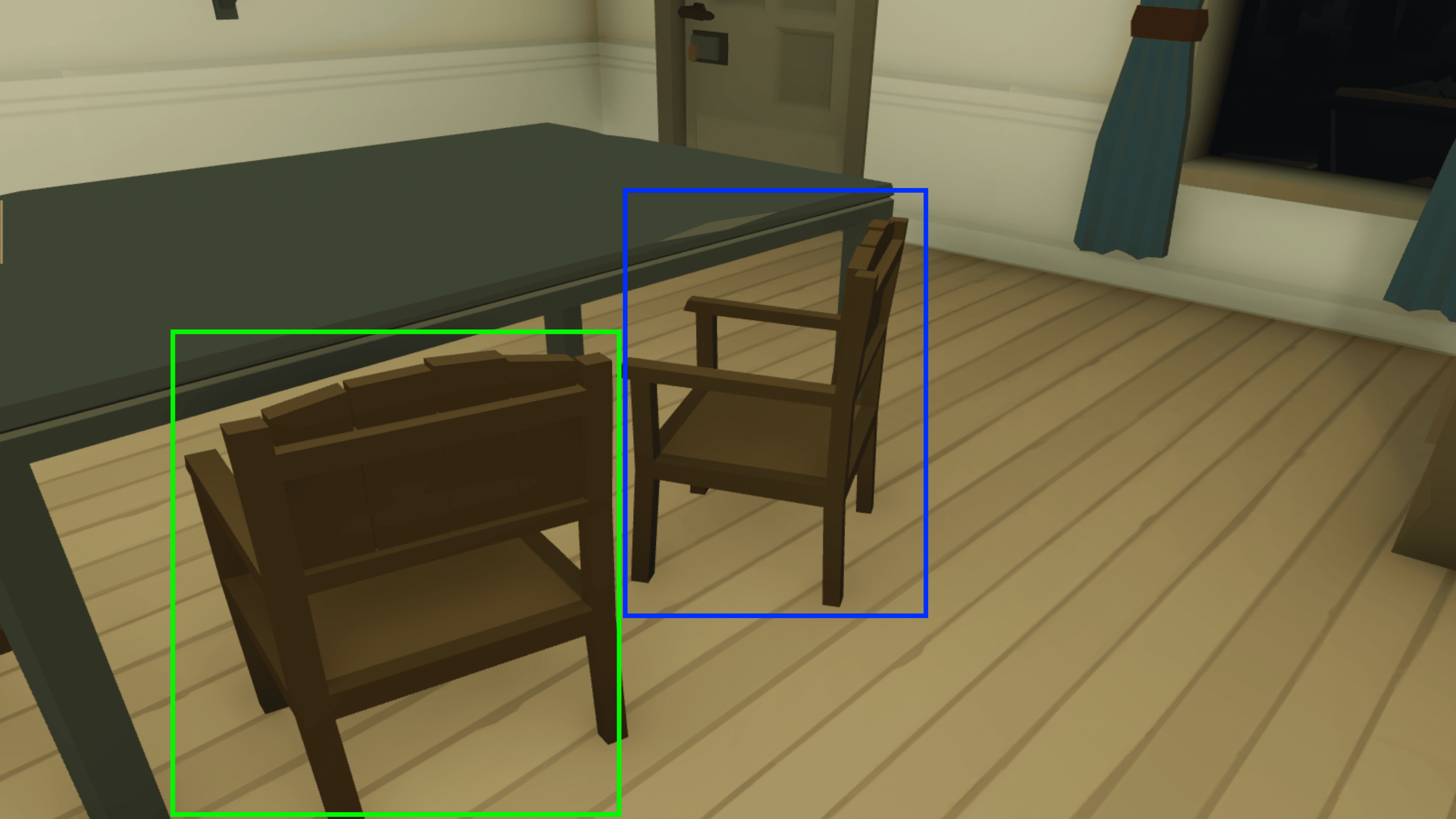}
    \end{subfigure}
    \caption{Example of FOV Pair for Same Entity Determination.}  
    \label{fig:study5_same_object}
\end{figure}

In this controlled study, 
we select the core features from RQ2 as context options.  
Prompts are then designed to ask the LLM to determine if two entities are the same target under
the consideration of different combinations of these contextual features.
Figure~\ref{fig:study5_same_object} shows an annotated FOV, where we manually label the entities with different color boxes to indicate the groundtruth.
In particular, the entities in the red and the green boxes are the same among these two FOVs.

\textbf{Prompt Design.}
Table~\ref{tab:object_determination} shows the details of the designed prompt.
This prompt is derived from the optimal prompt of RQ2. Specifically, the first part assigns the task with detailed requirements and context, while the second part refines LLM's response through chain-of-thought reviews. 
Note that we parameterize the \{box1\_color\}, \{box2\_color\} to automatically compare the entities in \textit{red-and-green} or \textit{red-and-blue}, and also with different \{feature\} combinations.

\begin{table}[!ht]
\centering
\caption{Detailed Description and Structured Data Rules for Image Analysis}
\begin{tabularx}{\linewidth}{|l|X|}
\hline
\textbf{Prompt} & \textbf{Description} \\
\hline
Designed Prompt Part 1 & \makecell[Xt]{Your task is to analyze two images and determine if the entities in the specified regions are the same entity. \\
For Image 1, focus on the entity within the \{box1\_color\} box. \\
For Image 2, focus on the entity within the \{box2\_color\} box. \\
Give priority to the entity's \{feature\} \\
Please provide: \\
1. Whether they are the same entity (true/false) \\
2. Your confidence score (0.0 to 1.0) \\
3. Your reasoning for your decision } \\
\hline
Designed Prompt Part 2 & \makecell[Xt]{Please review your comparison carefully. Consider: \\
1. Are there any subtle details you might have missed? \\
2. Could any entity differences affect your judgment? \\
3. Are there any unique identifying features that definitively prove or disprove it's the same entity? \\
Provide an updated analysis if needed. } \\
\hline
\end{tabularx}
\label{tab:object_determination}
\end{table}

\subsubsection{Evaluation Metrics}
With three selected contexts, we eventually prepare eight different combinations for each prompt: one for no context, one for all three contexts, three for two contexts, and three for one context. 
Besides outputting the true or false answer, we also ask the model to generate the confidence score (from $0.0$ to $1.0$) for its decision and a paragraph of reasoning to explain the decision.
Finally, we calculate the Precision, Recall, and F1-Score to evaluate the overall performance of using each context combination. Note that, we count the correct determination rate as follows: 

\begin{itemize}
\item True Positives (TP): The same entity that has been determined as the same target across multiple FOVs.
\item False Positives (FP): The same entity that has been determined as different targets across multiple FOVs.
\item True Negative (TN): The different entities that have been determined as different targets across multiple FOVs.
\item False Negative (FN): The different entities that have been determined as the same target across multiple FOVs.
\end{itemize}
\section{Evaluation} \label{sec:evaluation}

\textbf{Experiment Setup.}
The details of FOV creation, device, and application selection have been introduced in Section \ref{sec:data_construction}.
In summary, we select a simulated VR game titled \textit{The Break-In} and focus on a virtual house with different rooms. We use the Meta Quest 2 headset to capture screenshots representing a merged vision of the left and right eyes. The prompt designs and examples are discussed in Section~\ref{sec:method}. We select GPT-4o as the Large Language Model (LLM) for this study due to its practical usability and prompt-based functionality. In particular, we use the gpt-4o-2024-08-06 API with structured output \cite{openai_docs_structured_outputs} for all RQs. Note that, we use default model parameters, specifically, we set ``frequency\_penalty: $0$; presence\_penalty: $0$; temperature: $1$; and top\_p: $1$.''  For RQ4, we introduce ChatGPT-4o \cite{chatgpt} for image labeling tasks, as the APIs do not have this functionality.

In the following five sections, we present our findings from five designed studies. These include detailed statistical analyses, insights from manual observations, and in-depth reasoning discussions.

\subsection{Prompt Engineering for Efficient Entity Detection}

\begin{tcolorbox}[colback=gray!5!white,colframe=black!75!black]
  \textbf{RQ1}:  Can prompt engineering be used to enhance the accuracy of entity detection from an FOV in VR? \\
  \textit{Findings:} The basic prompt achieves $41.67\%$ ($90$ out of $216$) CDR. With prompt engineering, specifically chain-of-thought, the revised prompt increases the CDR to $71.30\%$ ($154$ out of $216$) and also maintains a relatively low EDR of $6.01\%$ ($13$ out of $216$).
\end{tcolorbox}

\begin{figure}[h]
    \centering
    \includegraphics[width=0.8\linewidth]{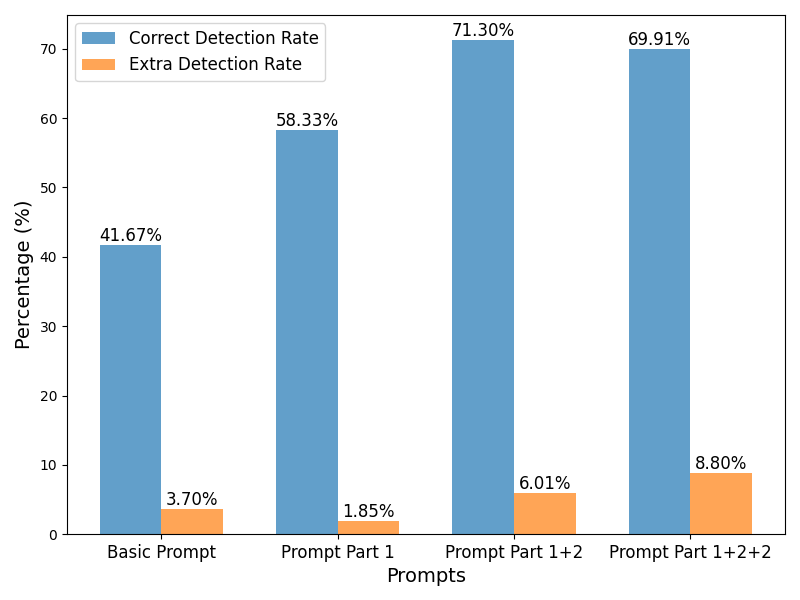}
    \caption{Detection Results between Different Prompts}
    \label{fig:report1_prompt}
\end{figure}

The detection results for different prompt combinations are shown in Figure~\ref{fig:report1_prompt}. We can see that the basic prompt, without any additional context or requirements, yields a $41.67\%$ Correct Detection Rate (CDR). When we apply Designed Prompt Part 1, which includes detailed requirements, the CDR increases to $58.33\%$ ($126$ out of $216$). Adding one iteration of Part 2, which asks GPT-4o to refine the results, the CDR rises further to $71.30\%$. However, when we apply two iterations of Part 2 for further refinement, the CDR slightly decreases to $69.91\%$ ($151$ out of $216$).
Additionally, we introduce a second metric, the Extra Detection Rate (EDR), to measure the proportion of incorrectly detected entities. When two iterations of Part 2 are applied, the EDR increases significantly from $6.01\%$ to $8.79\%$ ($19$ out of $216$), indicating that over-refining introduces more errors than it corrects.
In summary, a structured prompt with detailed requirements and a single round of self-correction achieves a higher CDR while maintaining a lower EDR.

\begin{table}[ht]
    \centering
    \caption{Prompt 1+2 Performance on Three Complexity Levels}
    \label{tab:best_prompt_analysis}
    \begin{tabular}{|l|c|c|c|c|}
        \hline
        \textbf{Metrics} & \textbf{Easy (\%)} & \textbf{Medium (\%)} & \textbf{Hard (\%)} & \textbf{Overall (\%)} \\ \hline
        Correct Detection Rate & 97.22 & 87.88 & 53.51 & 71.30 \\ \hline
        Extra Detection Rate & 16.66 & 7.57 & 1.75 & 6.01 \\
        \hline
    \end{tabular}
\end{table}

\begin{figure}
    \centering
    \begin{subfigure}{0.48\textwidth}
        \includegraphics[width=\linewidth]{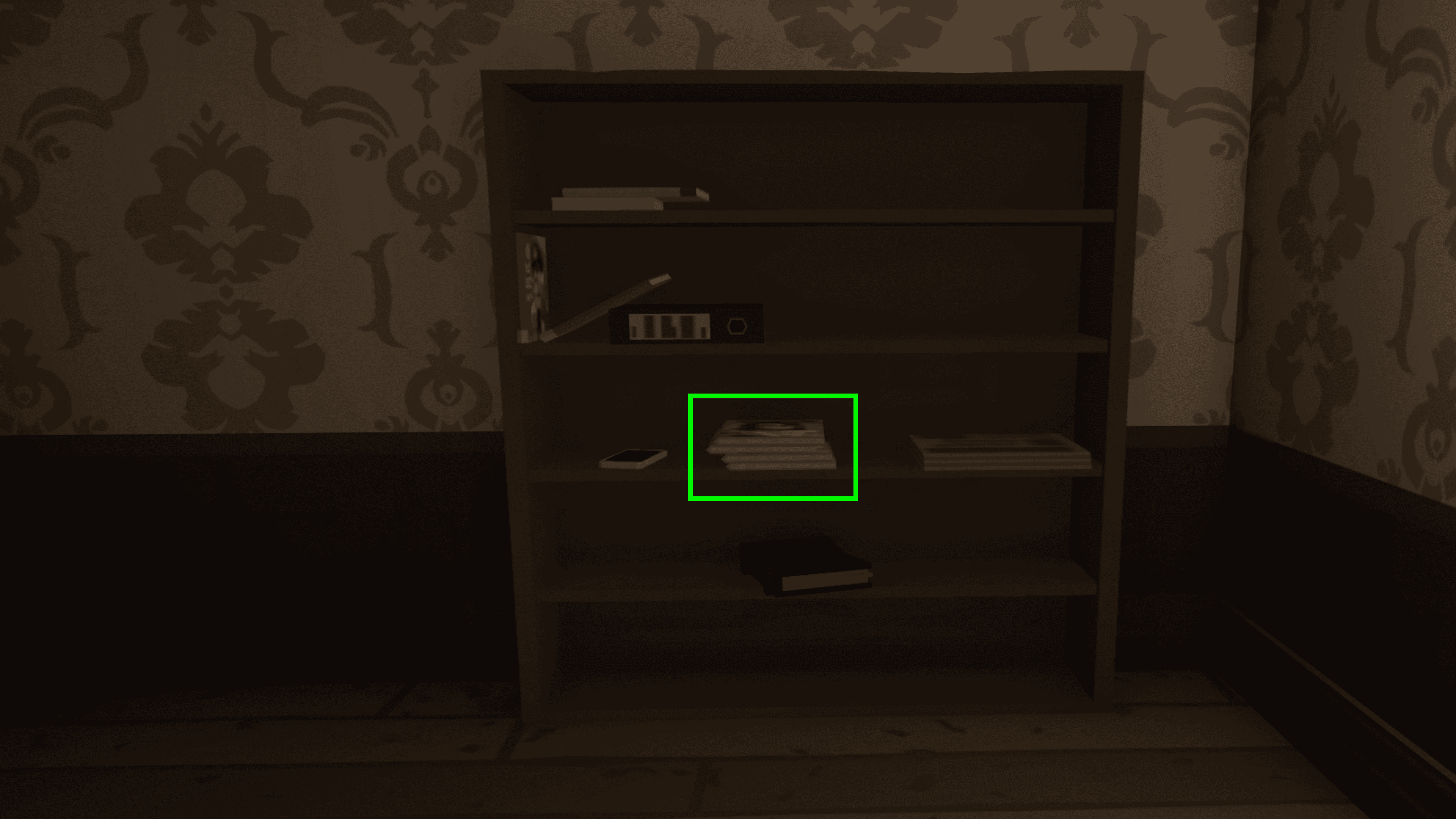}
        \caption{A Stack of Magazines}
    \end{subfigure}
    \begin{subfigure}{0.48\textwidth}
        \includegraphics[width=\linewidth]{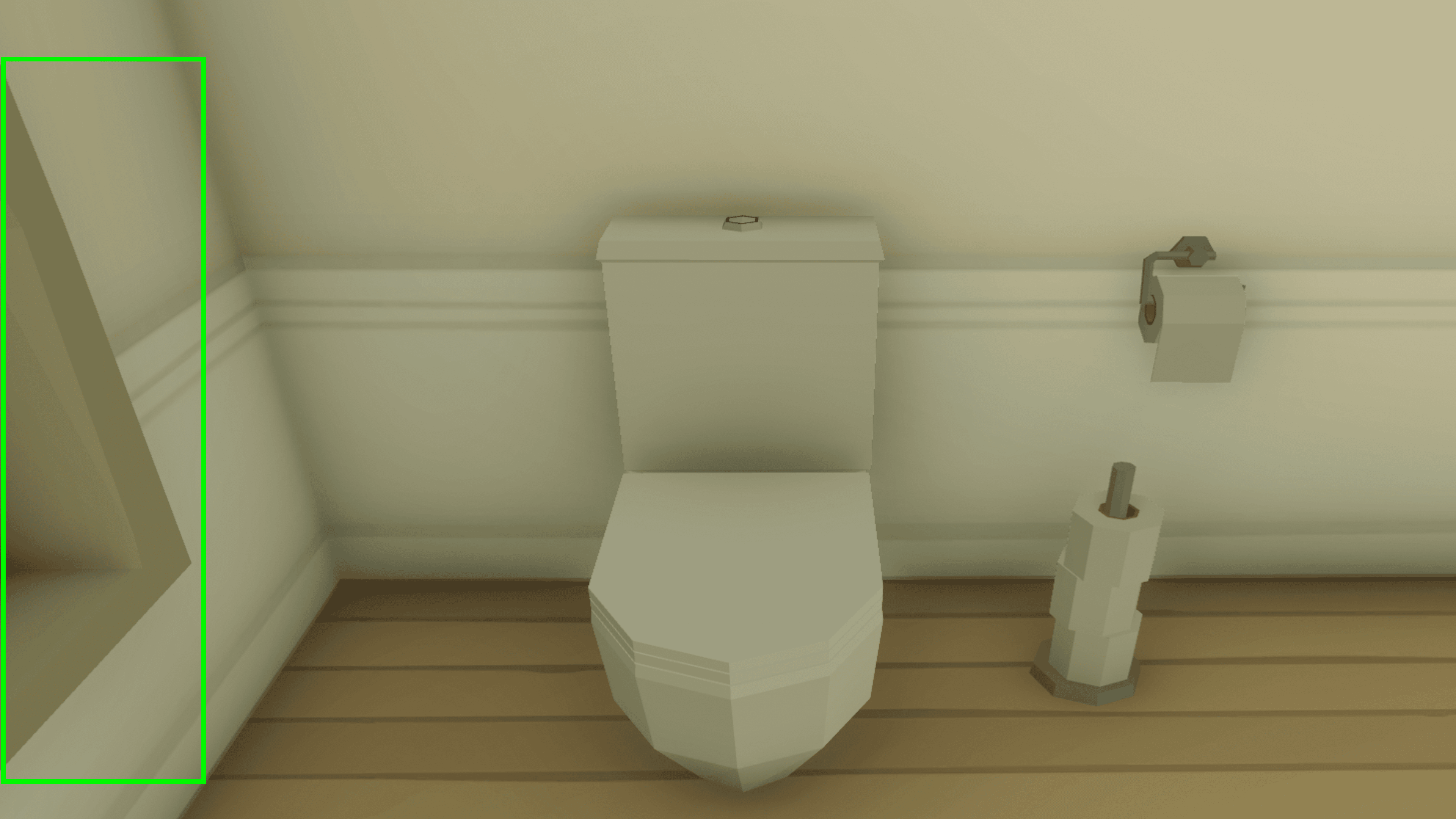}
        \caption{Extra Detected Wall light}
    \end{subfigure}
    \caption{Examples of Entity Identification Mistakes}
    \label{fig:report1_miss_detect}
\end{figure}

We further compare the performance of the optimal prompt combination across different scene complexity levels. Table~\ref{tab:best_prompt_analysis} presents the CDR and EDR for three levels of complexity. 
We find that the Easy level achieves the highest CDR, at $97.22\%$ ($35$ out of $36$), due to fewer entities in the scene. 
In contrast, the Hard level exhibits the lowest EDR of $1.75\%$ ($2$ out of $114$).
This is because the scene contains more entities, making it less likely to over-predict extra items.
Additionally, a notable CDR difference is observed across the Easy, Medium, and Hard levels. Through qualitative analysis of the missed detections, we find that GPT-4o tends to miss entities when they are grouped together. For instance, instead of detecting four individual books, GPT outputs a result of a “stack of papers,” even when the prompt specifies entity separation. This is illustrated in Figure~\ref{fig:report1_miss_detect} (a).
As scene complexity increases, entities are often placed together, which may explain the observed decrease in CDR in Table~\ref{tab:best_prompt_analysis}. 
On the contrary, the EDR keeps decreasing from easy to hard scenes, and this also explains that over-correction will bring more errors when GPT-4o has already achieved a higher detection rate. As shown in Figure~\ref{fig:report1_miss_detect} (b), an extra wall light is detected during refining.
This limitation underscores the need for further improvement and fine-tuning of the LLM to better handle the detection of grouped entities.

\subsection{Accuracy of Features Identification}

\begin{tcolorbox}[colback=gray!5!white,colframe=black!75!black]
  \textbf{RQ2}: What are the core features for identifying test entities in FOV, and how accurately can LLM describe them? \\
  \textit{Findings:} Among all the contextualized five features, we identify color, placement, and shape as three core features to identify entities in the VR environment, with the description accuracies of 
  $94.80\%$ ($146$ out of $154$), $95.45\%$ ($147$ out of $154$), and $96.10\%$ ($148$ out of $154$), respectively.
\end{tcolorbox}

As discussed in Section~\ref{subsec:study2}, we are going to measure the LLM identification accuracy on each entity of its five contextualized features: \textit{color}, \textit{placement} (location and orientation), \textit{lighting}, \textit{texture and materials}, and \textit{shape}.
We first ask GPT-4o to describe each feature for the detected entities and then compare it with the groundtruth. Three annotators then vote to decide whether the generated description matches the groundtruth data or not.
The final results are shown in Table~\ref{tab:report2_FIR}.
GPT-4o demonstrates strong performance in detecting all features in general with above $90\%$ accuracy over $216$ groundtruth labels.
In particular, feature \textit{lighting} performs the best with a $98.05\%$ rate ($151$ out of $154$), while feature \textit{texture and materials} has the lowest accuracy rate of $90.25\%$ ($139$ out of $154$).

\begin{table}[h]
    \centering
    \caption{Features Identification Rate}
    \label{tab:report2_FIR}
    \begin{tabular}{|p{3cm}|p{2cm}|}
        \hline
        {\textbf{Metrics}} & \textbf{Overall (\%)} \\ \hline
        Color & 94.80 \\ \hline
        Placement  & 95.45 \\ \hline
        Lighting  & 98.05 \\ \hline
        Texture \& Material  & 90.25 \\ \hline
        Shape  & 96.10 \\ \hline     
    \end{tabular}
\end{table}

\begin{table}[h]
    \centering
    \caption{Features Identification Rate under Different Lighting}
    \label{tab:report2_lighting}
    \begin{tabular}{|l|c|c|c|}
        \hline
        {\textbf{Metrics}} & \textbf{Light on (\%)} & \textbf{Light off (\%)} & \textbf{Flash Light (\%)} \\ \hline
        Color & 96.36 & 93.02 & 94.64  \\ \hline
        Placement & 96.36 & 93.02 & 96.42  \\ \hline
        Texture \& Material & 90.90 & 90.69 & 89.28  \\ \hline
        Shape & 96.36 & 95.34 & 96.42  \\ \hline      
        \end{tabular}
\end{table}

\begin{figure}[h]
    \centering
    \hfill
    \begin{subfigure}[b]{0.48\textwidth}
        \includegraphics[width=\linewidth]{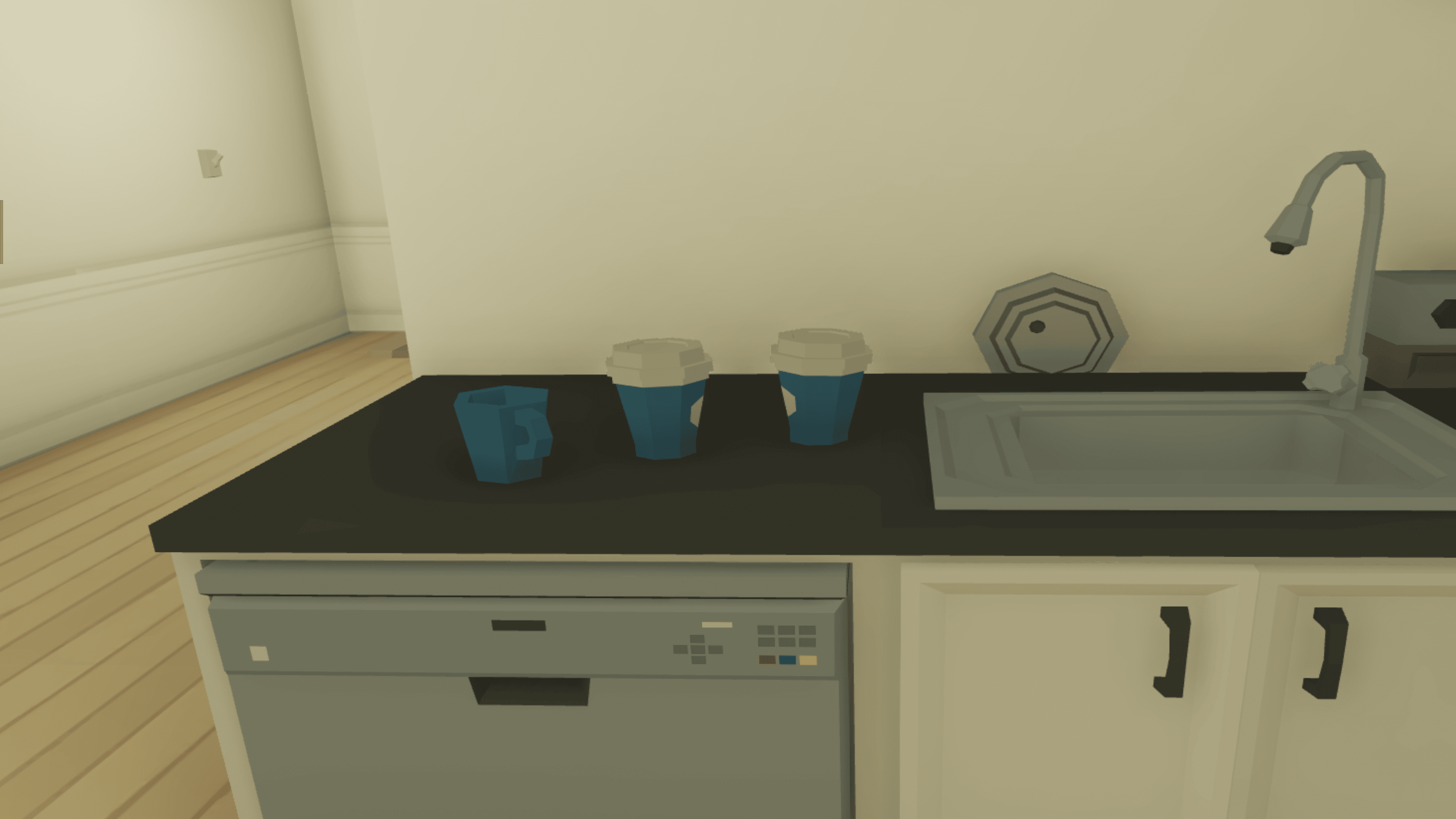}
    \end{subfigure}
    \hfill
    \begin{subfigure}[b]{0.48\textwidth}
        \includegraphics[width=\linewidth]{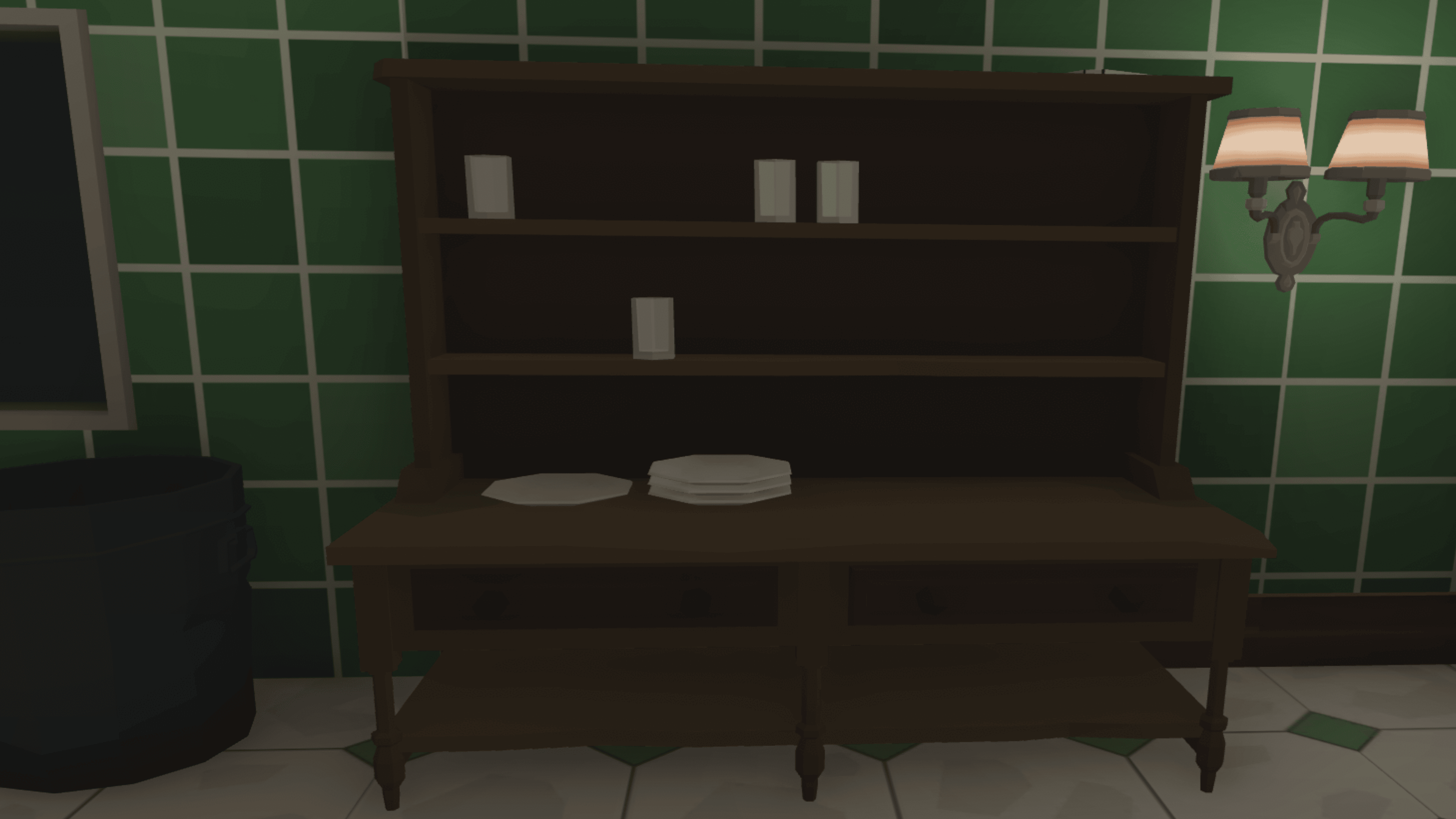}
    \end{subfigure}
    
    \caption{Examples of Texture and Material}
    \label{fig:scene_texture_material}
\end{figure}

We further investigate the reasons behind GPT-4o’s lower accuracy in recognizing texture and material. We find that the cartoonish and non-photorealistic game design style simplifies many textures, making it challenging for the LLM to identify them accurately. Additionally, these simplified representations also pose difficulties for human annotators, who struggle to differentiate textures from one another.
For example, in Figure~\ref{fig:scene_texture_material}, the blue cup with a white lid on the countertop is difficult to classify as either a plastic, ceramic, or paper cup. Similarly, in the right screenshot, it is unclear whether the bottles on the top shelf have a glass, ceramic, or plastic texture.
To avoid further ambiguity, we exclude texture and material as the key features in entity identification for the remainder of our study.

\begin{figure}[h]
    \centering
    \begin{subfigure}[b]{0.32\textwidth}
        \includegraphics[width=\linewidth]{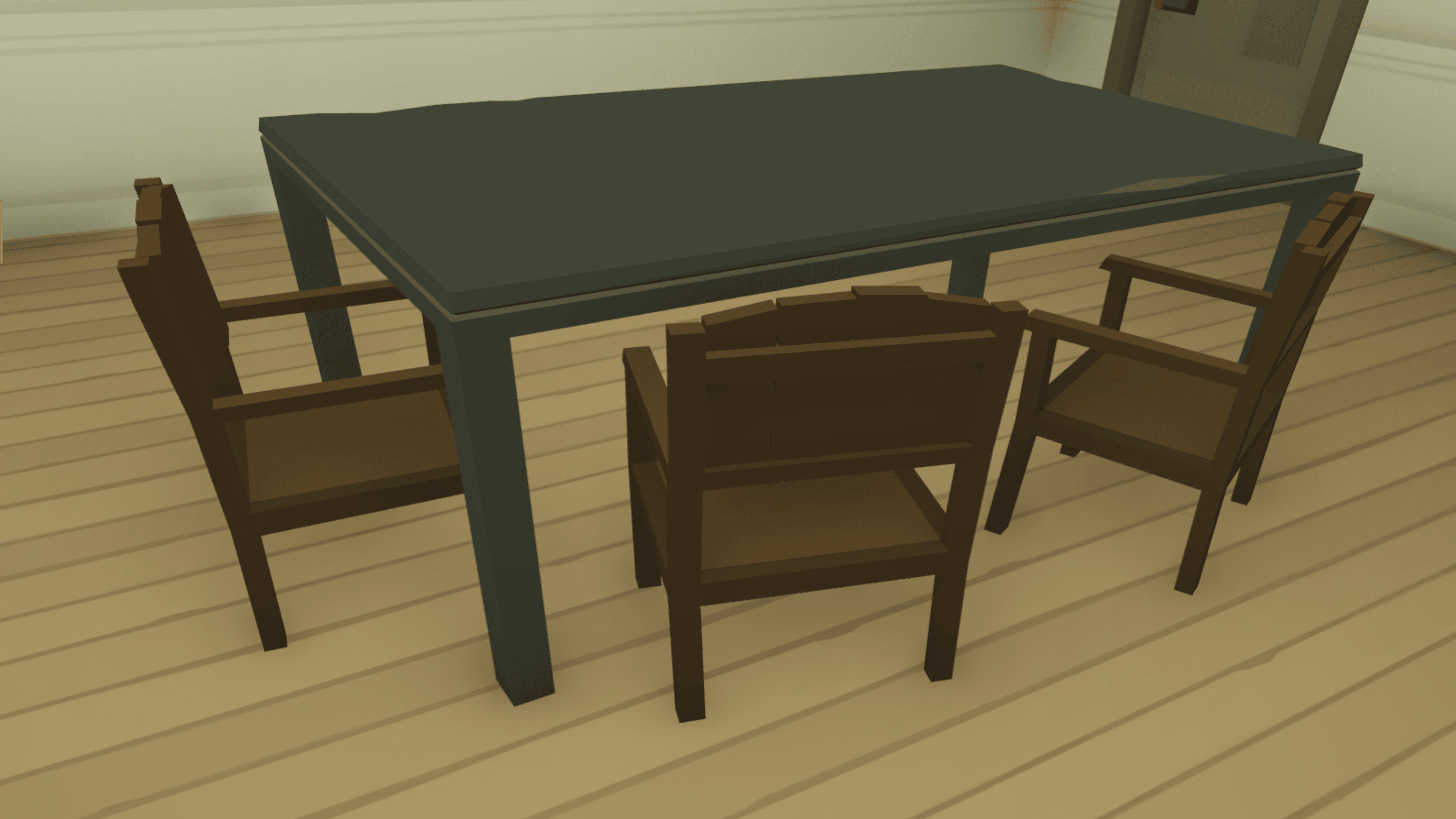}
        \caption{Light on}
    \end{subfigure}
    \hfill
    \begin{subfigure}[b]{0.32\textwidth}
        \includegraphics[width=\linewidth]{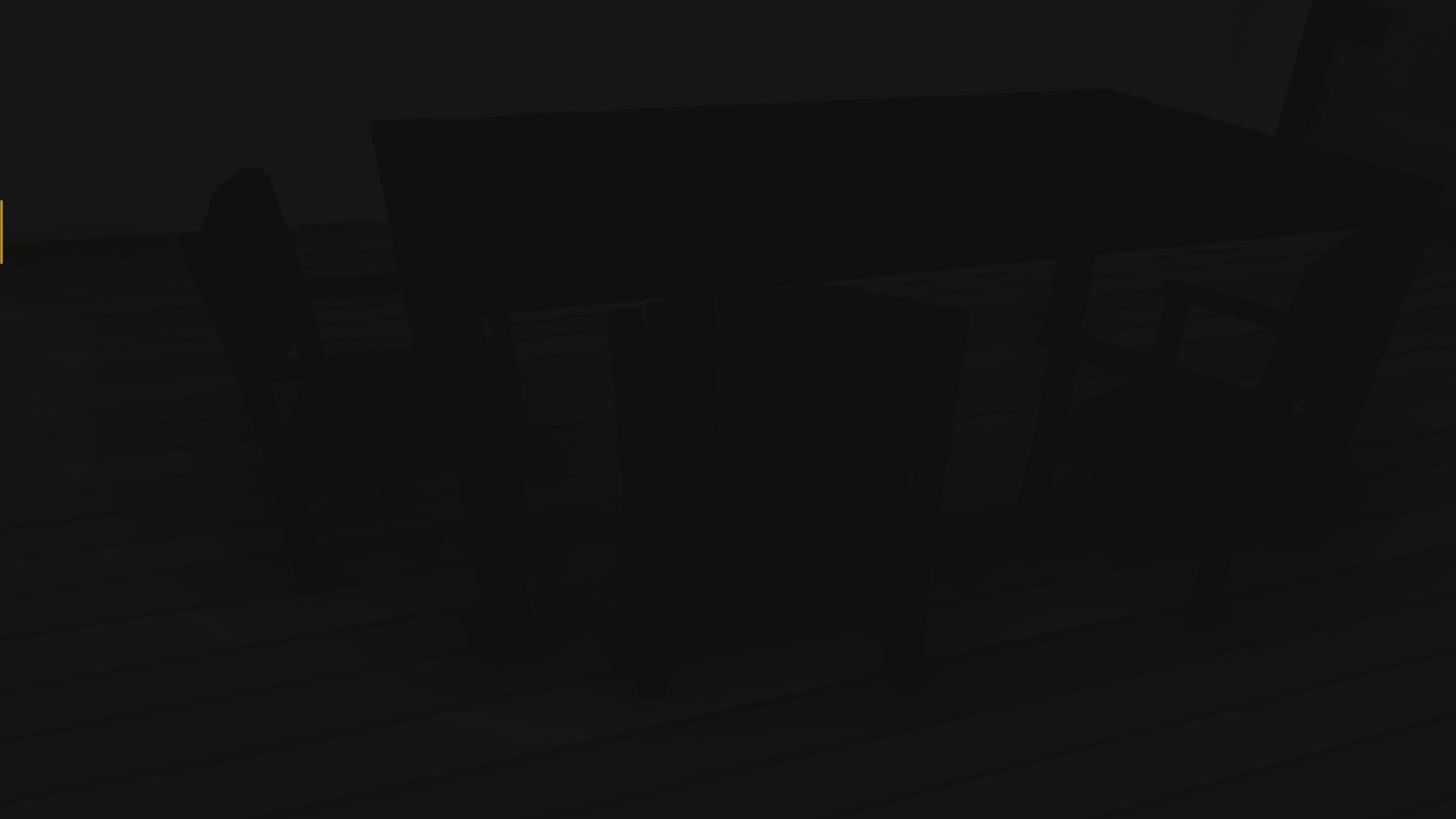}
        \caption{Light off}
    \end{subfigure}
    \hfill
    \begin{subfigure}[b]{0.32\textwidth}
        \includegraphics[width=\linewidth]{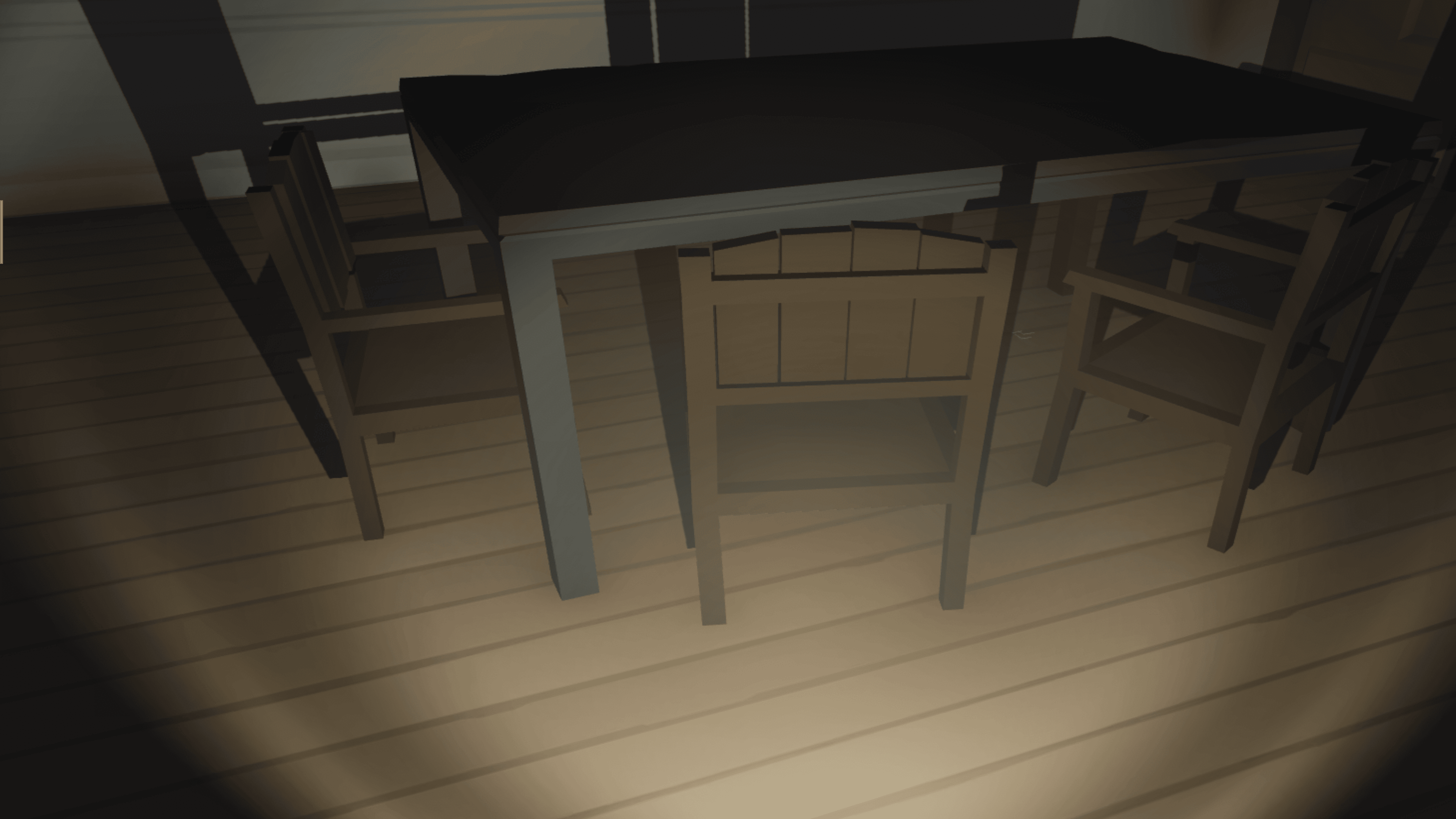}
        \caption{Flash Light}
    \end{subfigure}
    
    \caption{Examples of Lighting at the Same Scene}
    \label{fig:scene_different_lighting}
\end{figure}

While lighting shows the highest detection success rate, further investigation (shown in Table~\ref{tab:report2_lighting}) reveals that feature identification accuracy remains consistent across varying lighting conditions within the same scene. This is reasonable, as GPT-4o can ``see'' details at the pixel level that the human eye might miss. Additionally, when the lights are off (shown in Figure~\ref{fig:scene_different_lighting} (b)), human annotators find it challenging to verify the accuracy of feature identification (the hardness level depends on factors including monitor models and monitor settings). However, GPT-4o is not impacted by these human-affecting factors, yielding more stable and more accurate results.
These findings suggest that lighting has no significant impact on GPT-4o's ability to recognize entities. Therefore, we exclude lighting as a key feature in our analysis.

Based on these observations and discussions, the three most critical features for accurate entity identification are color, placement, and shape, which form the foundation for GPT-4o's entity detection capabilities in VR environments.

\subsection{Accuracy of Scene and Spatial Analysis}

\begin{tcolorbox}[colback=gray!5!white,colframe=black!75!black]
  \textbf{RQ3}: How accurate is LLM in scene recognition and spatial understanding through FOV analysis? \\
  \textit{Findings:} GPT-4o correctly identifies $85.18\%$ ($23$ out of $27$) of the scene, and with the designed structured output prompts, it accurately describes $92.86\%$ of the spatial relationships between the user’s view and the target entities.
\end{tcolorbox}

Out of $27$ scene screenshots, GPT-4o successfully recognizes $85.18\%$ of them.
After reviewing the misclassified results, we find that the subtle differences between GPT-4o's answers and the groundtruth labels are the main reasons for the error. GPT-4o often uses similar or closely related concepts to describe the scene. 
For example, sometimes, GPT-4o recognizes a room with a crib and toys as a ``children's room,'' while other times, it describes the same room as a ``nursery room.'' This is unstable and does not always match the groundtruth label of ``nursery room.''
Since we do not provide predefined terms for scene categorization and rely solely on GPT-4o’s general knowledge to identify and summarize the scene in a short phrase, such subtle variations in the final output are reasonable. However, these minor deviations will hinder further automated testing classification. Therefore, it is necessary to establish an additional connection between the model’s output and the groundtruth labels in the future exploration design.

As discussed in Section~\ref{subsec:study3}, we first create a basic prompt to collect horizontal, vertical, and depth information between the user’s FOV and the target entity. However, GPT-4o’s response is inadequate due to missing data points. To address this,
we design prompts to collect structured outputs of the three spatial relationships for each entity. 
After applying the newly designed prompts, GPT-4o can generate information for all three directions.
Table~\ref{tab:report3_example} shows the number of reported data before and after we adopt the structured output.
We can see that among the nine examples from three different complexity levels, only Med 1 can generate all three direction data points, but the value remains unusable for quantitative comparison.
After applying the structured output design, all the returned data points contain three dimensions and can be compared quantitatively.

\begin{table}[h]
   \centering
   \caption{Comparison of Structured and Unstructured Prompts Output}
      \begin{tabular}{|c|c|c|c|c|}
      \hline
      \textbf{Entity} & \textbf{Unstructed Output} & \textbf{Dim} & \textbf{Structed Output} & \textbf{Dim} \\ \hline
      Easy 1 & right & 1 & right; down; near & 3 \\ \hline
      Easy 2 & center & 1 & center; center; mid-range & 3 \\ \hline
      Easy 3 & center left on countertop & 2 & left; up; near & 3 \\ \hline
            
      Med 1 & center; on table; far right & 3 & right; down; mid-range & 3 \\ \hline
      Med 2 & on counter; center left & 2 & left; center; mid-range & 3 \\ \hline
      Med 3 & in front of user & 1 & left; down; mid-range & 3 \\ \hline
      
      Hard 1 & around the table & 1 & left; center; mid-range & 3 \\ \hline
      Hard 2 & on the middle shelf of Cabinet A & 1 & middle; up; mid-range & 3 \\ \hline
      Hard 3 & on the top shelf of the bookshelf & 1 & right; up; mid-range & 3 \\ \hline

      \end{tabular}
      \label{tab:report3_example}
\end{table}

Table~\ref{tab:report3_spatial} shows the quantitative evaluation of GPT-4o’s response after requesting structured outputs to collect its understanding of spatial relationships.
The horizontal relationship achieves the highest accuracy, with an overall rate of $89.61\%$, because this concept is less complex and easier to detect. Depth follows with the second-highest accuracy of $81.82\%$, while the vertical relationship has the lowest accuracy of $72.90\%$.
After further investigation, we identify a potential cause for this misalignment, which is the comprehension difference between the 2D image and the 3D scene.
Specifically, for the groundtruth, we follow the image analysis method in photography, where the entity’s vertical location is determined by dividing the FOV screenshot into three equal parts. 
These three parts are mapped to the up, center, and down for the vertical locations.
However, GPT-4o analyzes the vertical relationship by considering the user’s vertical perspective.
For example, in Figure~\ref{fig:study2_feature_identification_new}, 
the chair in the middle is labeled as ``center'' in the groundtruth because it is positioned centrally in the image. 
GPT-4o, however, outputs ``down,'' likely because the chair is placed on the ground, making its vertical position lower than the user’s view.

\begin{table}[!t]
    \centering
    \caption{Accuracy of Spatial Relationship Detection across Different Scenes}
    \label{tab:report3_spatial}
    \begin{tabular}{|c|c|c|c|c|}
        \hline
        \textbf{} & \textbf{Easy Scene (\%)} & \textbf{Medium Scene (\%)} & \textbf{Hard Scene (\%)} & \textbf{Overall (\%)} \\ \hline
        Horizontal & 94.29 & 84.21 & 93.44 & 89.61 \\ \hline
        Vertical & 68.57 & 63.16 & 83.61 & 72.90 \\ \hline
        Depth & 85.71 & 80.70 & 81.97 & 81.82 \\ \hline
    \end{tabular}
\end{table}

\subsection{Capability of Labeling Identified Entity}

\begin{tcolorbox}[colback=gray!5!white,colframe=black!75!black]
  \textbf{RQ4}: Can LLM be used to label the identified entities in VR? \\
  \textit{Findings:} After employing four approaches that include image resizing and coordinates-based labeling, we, unfortunately, find that none of the entities have been correctly labeled in $27$ FOVs. 
\end{tcolorbox}

\begin{figure}[ht]
    \centering
    \includegraphics[width=\linewidth]{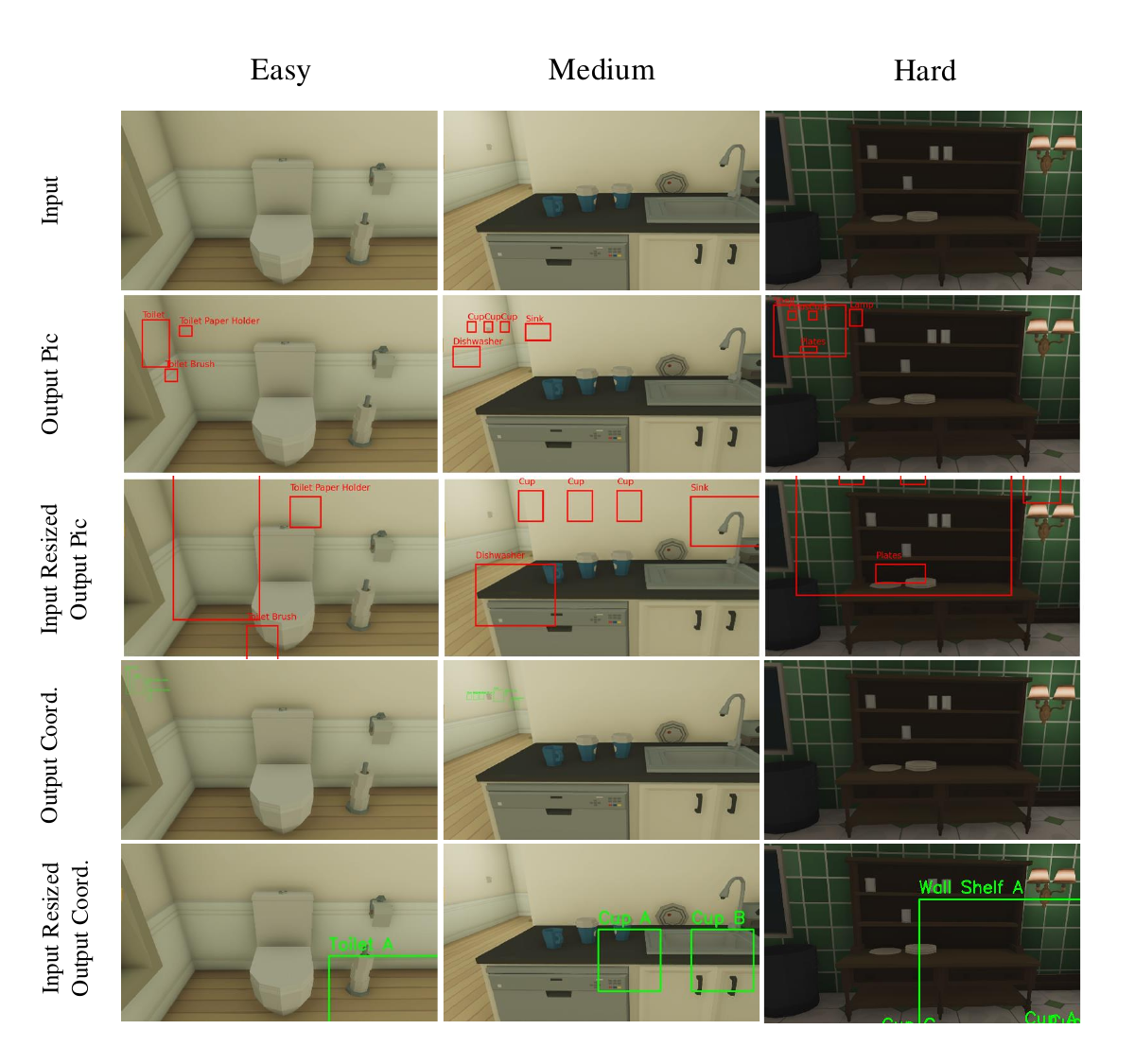}
    \caption{Comparison of outputs with and without resizing across models. Each column represents a different difficulty level (Easy, Medium, Hard), and each row shows the results from various workflows.}
    \label{fig:comparison_outputs}
\end{figure}

To comprehensively evaluate the impact of preprocessing and annotation methods on entity detection performance, we conduct four experimental setups emphasizing two distinct workflows: direct image annotation and coordinate-based detection.

As discussed in Section~\ref{subsec:study4}, in approach 1, raw images are fed into ChatGPT-4o, which directly returns annotated images with bounding boxes, serving as a baseline for evaluation. Similarly, in approach 2, images are preprocessed by adding padding to achieve square dimensions and are resized to $512\times512$ pixels before putting into ChatGPT-4o, because an image with a dimension exceeding $512\times512$ would otherwise be divided into smaller $512\times512$ tiles, or be compressed into a single $512\times512$ image before processing~\cite{openai_vision_size}. Such black-box resizing can disrupt GPT-4o’s ability to consider spatial relationships between entities, potentially leading to inaccuracy. 
We try to use user-resizing to ensure uniform input size for more consistent results. 
Note that, these user-resized images are then directly annotated by the model.

In contrast, Approach 3 and Approach 4 employ a coordinate-based strategy. Approach 3 uses raw images, with the model returning entity coordinates (e.g., top-left and bottom-right corners) in response to specific prompts. These coordinates are subsequently visualized locally using OpenCV to generate the bounding boxes. Approach 4 applies this same coordinate-based detection to padded and resized images. This approach of extracting coordinates is based on the observation that LLMs exhibit limited proficiency in directly generating annotated images with bounding boxes~\cite{ContextDET}, making coordinate-based outputs a more reliable alternative.

However, despite employing multiple methods, the experimental results indicate that 
GPT-4o and ChatGPT-4o are not capable of correctly labeling identified entities.
Figure~\ref{fig:comparison_outputs} shows examples of the labeling results.
We can observe that none of the bound boxes has correctly enclosed the target entity. Some of them are shifted, and some do not even show up due to the wrong coordinates.
These findings suggest that more advanced LLMs or alternative approaches (Discussed in Section~\ref{sec:discussion}) are required for entity labeling tasks.

\subsection{Accuracy of Identical Entity Determination}

\begin{tcolorbox}[colback=gray!5!white,colframe=black!75!black]
  \textbf{RQ5}: Can core features enhance the accuracy of determining the same entities in multiple FOVs? \\
  \textit{Findings:} The combination of the core features—color, shape, and placement—outperforms all other feature combinations in identifying identical entities, achieving the highest F1-score of $0.70$ across $108$ determinations.
\end{tcolorbox}

\begin{table}[ht]
\centering
\caption{Classification Performance Metrics under Different Feature Combinations}
\begin{tabular}{|l|c|c|c|c|c|c|c|}
\hline
\textbf{Feature} & \textbf{TP} & \textbf{FP} & \textbf{TN} & \textbf{FN} & \textbf{Precision} & \textbf{Recall} & \textbf{F1-Score} \\ \hline
                   None & 34 & 20 & 34 & 20 & 0.63 & 0.63 & 0.63 \\ \hline
                  Color & 35 & 19 & 35 & 19 & 0.65 & 0.65 & 0.65 \\ \hline
       Color, Placement & 30 & 25 & 29 & 24 & 0.55 & 0.56 & 0.55 \\ \hline
           Color, Shape & 39 & 28 & 26 & 15 & 0.58 & 0.72 & 0.64 \\ \hline
Color, Shape, Placement & 40 & 20 & 34 & 14 & \underline{0.67} & \underline{0.74} & \underline{0.70} \\ \hline
              Placement & 27 & 21 & 33 & 27 & 0.56 & 0.50 & 0.53 \\ \hline
                  Shape & 34 & 28 & 26 & 20 & 0.55 & 0.63 & 0.59 \\ \hline
       Shape, Placement & 31 & 24 & 30 & 23 & 0.56 & 0.57 & 0.57 \\ \hline
\end{tabular}

\label{tab:performance_metrics}
\end{table}

The results in Table \ref{tab:performance_metrics} highlight the performance of GPT-4o in detecting the identical entity using various combinations of key features that we identified from RQ2, including \textbf{color}, \textbf{shape}, and \textbf{placement}. 
Among all the combinations, the combination of \textit{color}, \textit{shape}, and \textit{placement} achieves the highest F1-score of $0.70$ with the highest Recall of $0.74$ and the highest Precision of $0.67$.
The results indicate that the most effective approach for determining identical entities across multiple user FOVs is to have a comprehensive analysis while prioritizing all three key features of the target entity.
These findings further prove that to accurately understand and identify an arbitrary entity among FOVs, the decision model should systematically consider all its key features. 

\begin{figure}
  \centering
  \includegraphics[width=0.9\linewidth]{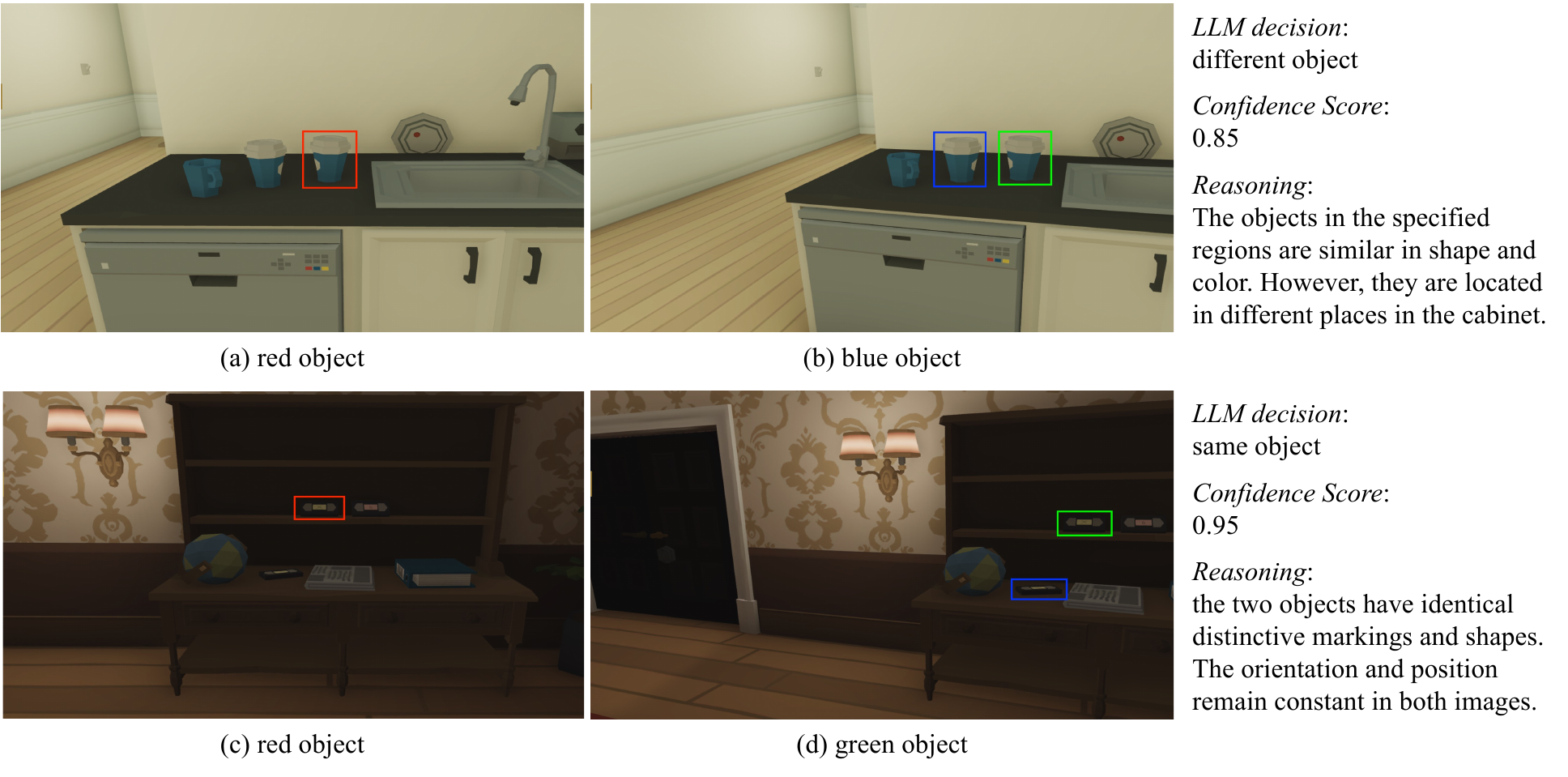}
  \caption{Examples of Correct Entity Determination with Reasoning}
  \label{fig:rq4_correct_examples}
\end{figure}

Figure~\ref{fig:rq4_correct_examples} shows two correct examples of GPT-4o determining the same and the different objects across multiple FOVs.
In the first example of comparing the red entity in (a) and the blue entity in (b), LLM
explains that the two entities are similar in shape and color. However, they have different locations in the cabinet and should consider different objects.
In the second example of comparing the red and green entities in (c) and (d), GPT-4o determines they are identical since the two objects have the same markings and shapes. Moreover, the orientation and position remain constant in both images.
We further examine the confidence scores and the generated reasoning for the failed cases and discuss in detail the reliability of GPT-4o's decisions.

\section{Discussion} \label{sec:discussion}

In this section, we review the limitations observed in the LLM, GPT-4o, during the experiments. 
Specifically, we will discuss the potential ways to improve entity labeling in RQ4, and the reliability of the entity determinations in RQ5. 
Additionally, we will explore some major implications of the findings from the five studies for future VR developers and VR testing researchers.

\subsection{Limitation and Improvement for LLM Labeling}

\begin{figure}
    \centering
    \begin{subfigure}{0.32\textwidth}
        \includegraphics[width=\linewidth]{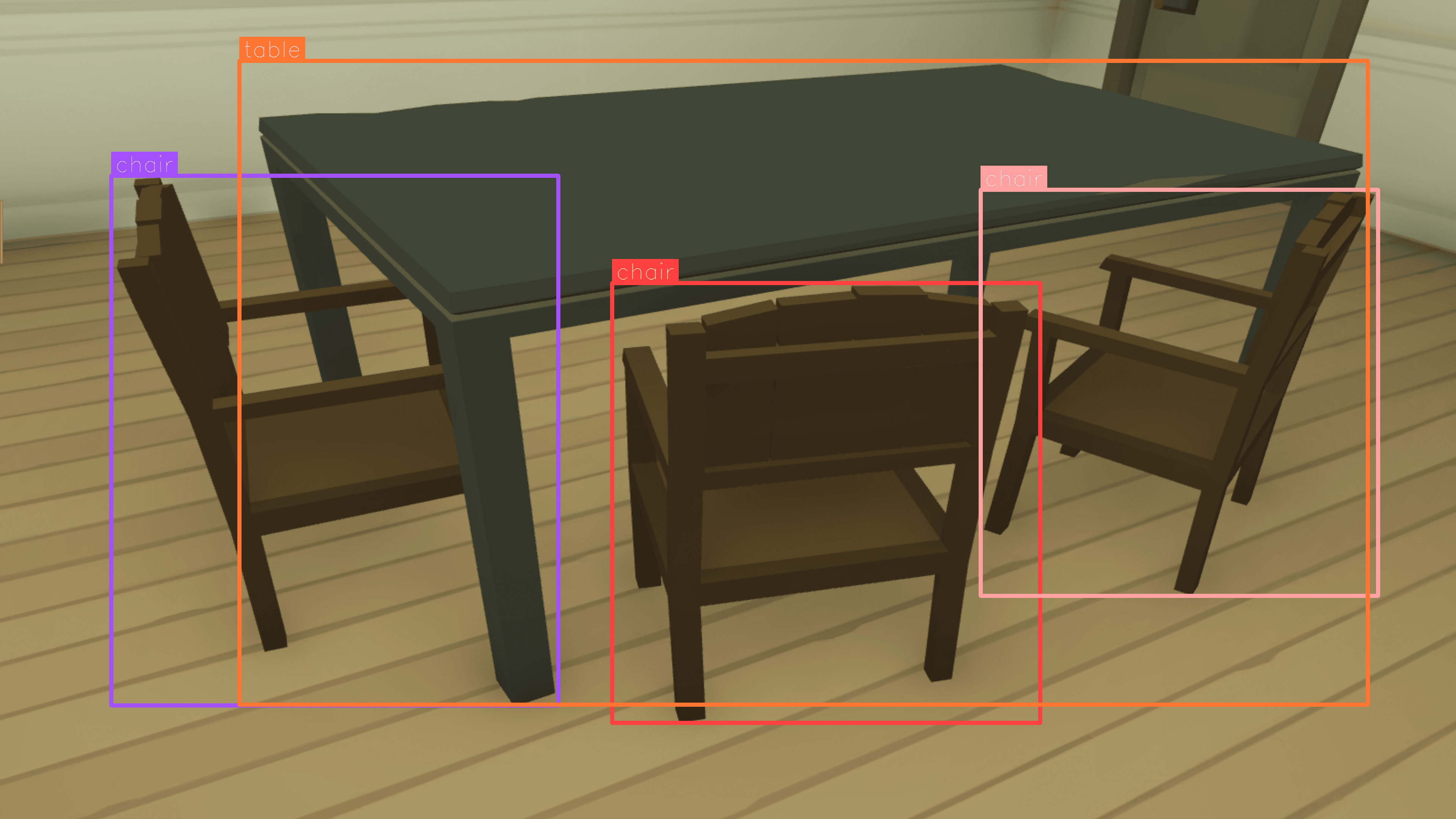}
        \caption{Easy}
    \end{subfigure}
    \hfill
    \begin{subfigure}{0.32\textwidth}
        \includegraphics[width=\linewidth]{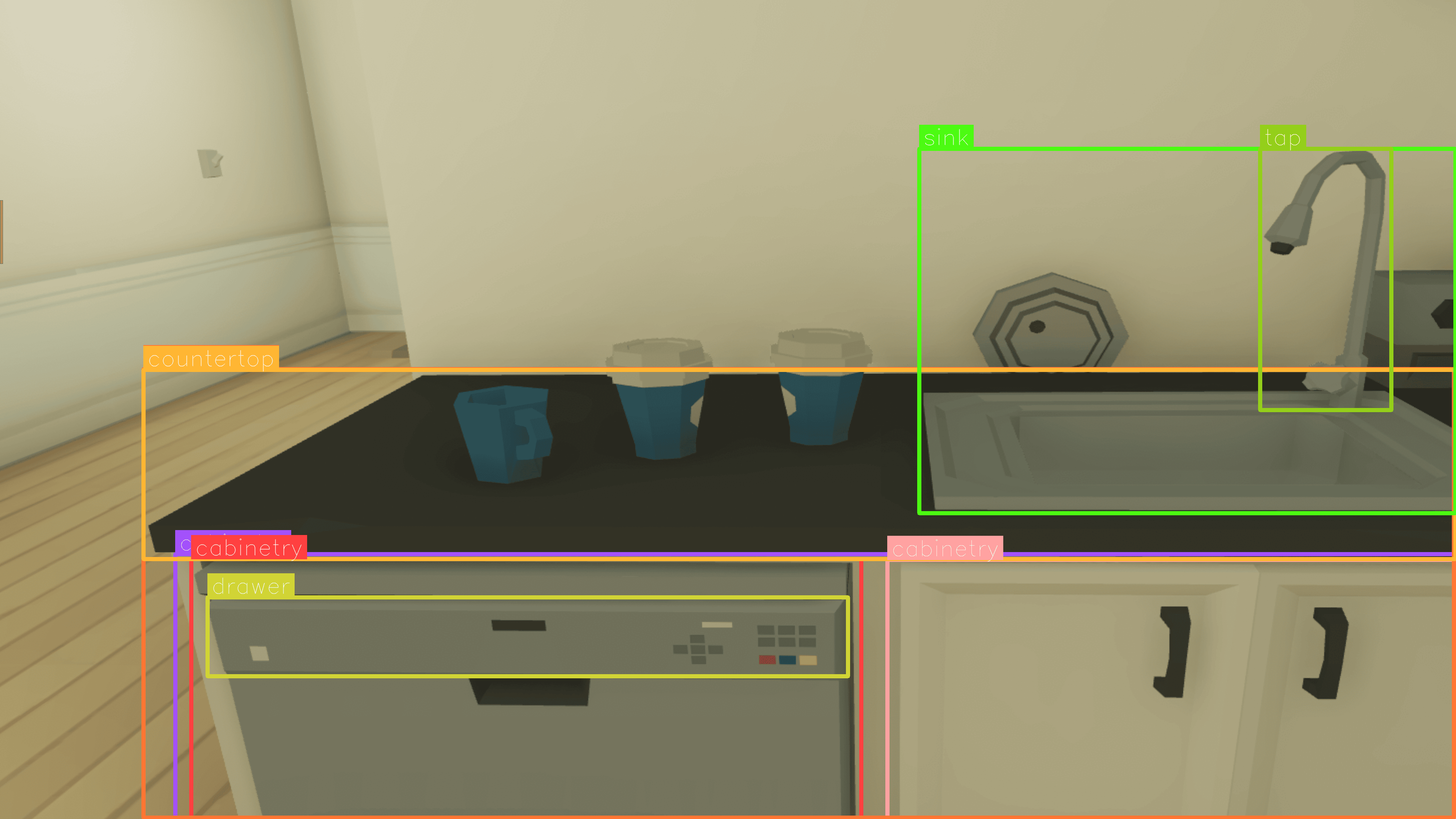}
        \caption{Medium}
    \end{subfigure}
    \hfill
    \begin{subfigure}{0.32\textwidth}
        \includegraphics[width=\linewidth]{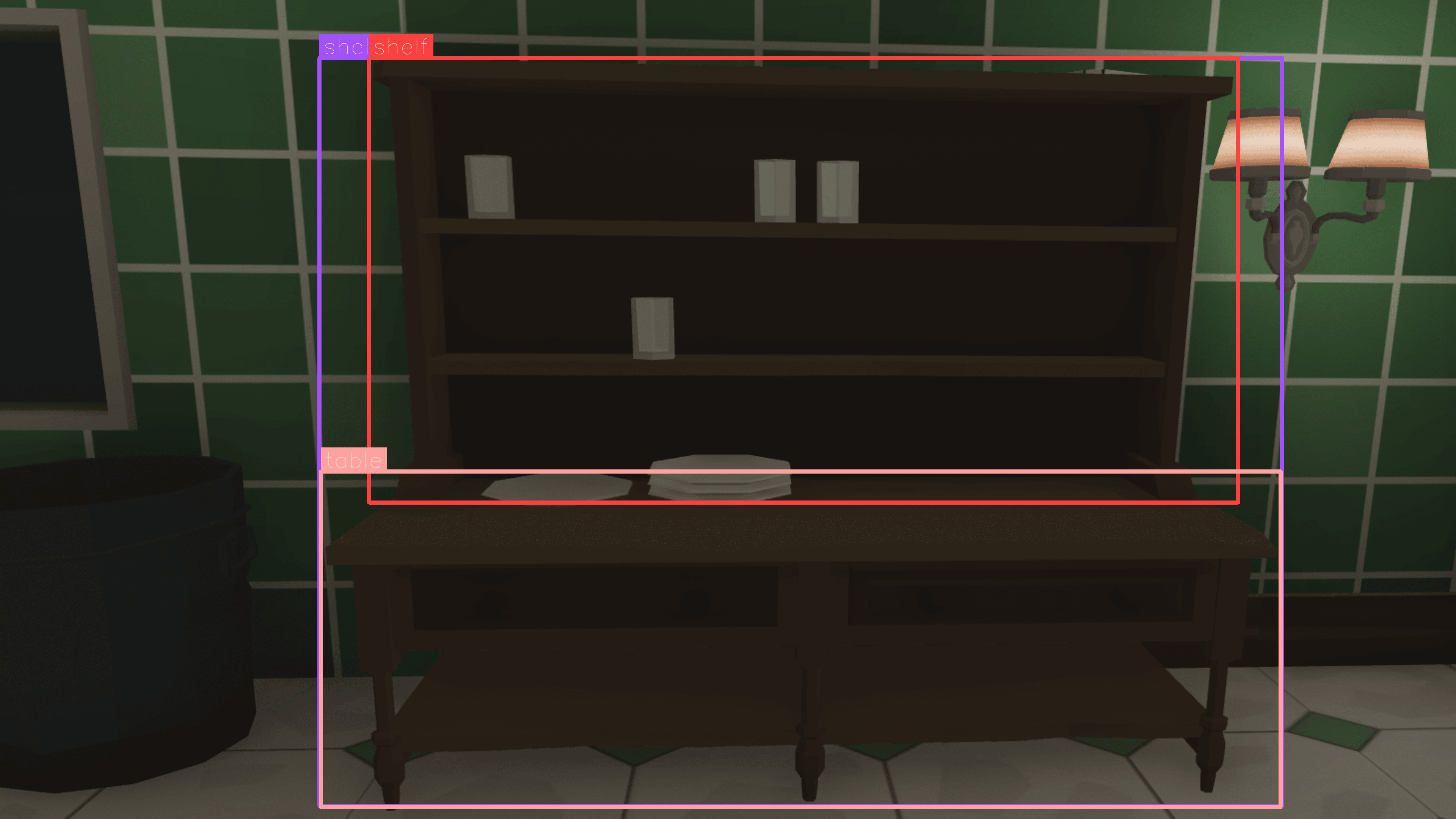}
        \caption{Hard}
    \end{subfigure}
    \caption{Result of Florence-2}
    \label{fig:florence_result}
\end{figure}

\begin{figure}
    \centering
    \begin{subfigure}[b]{0.32\textwidth}
        \includegraphics[width=\linewidth]{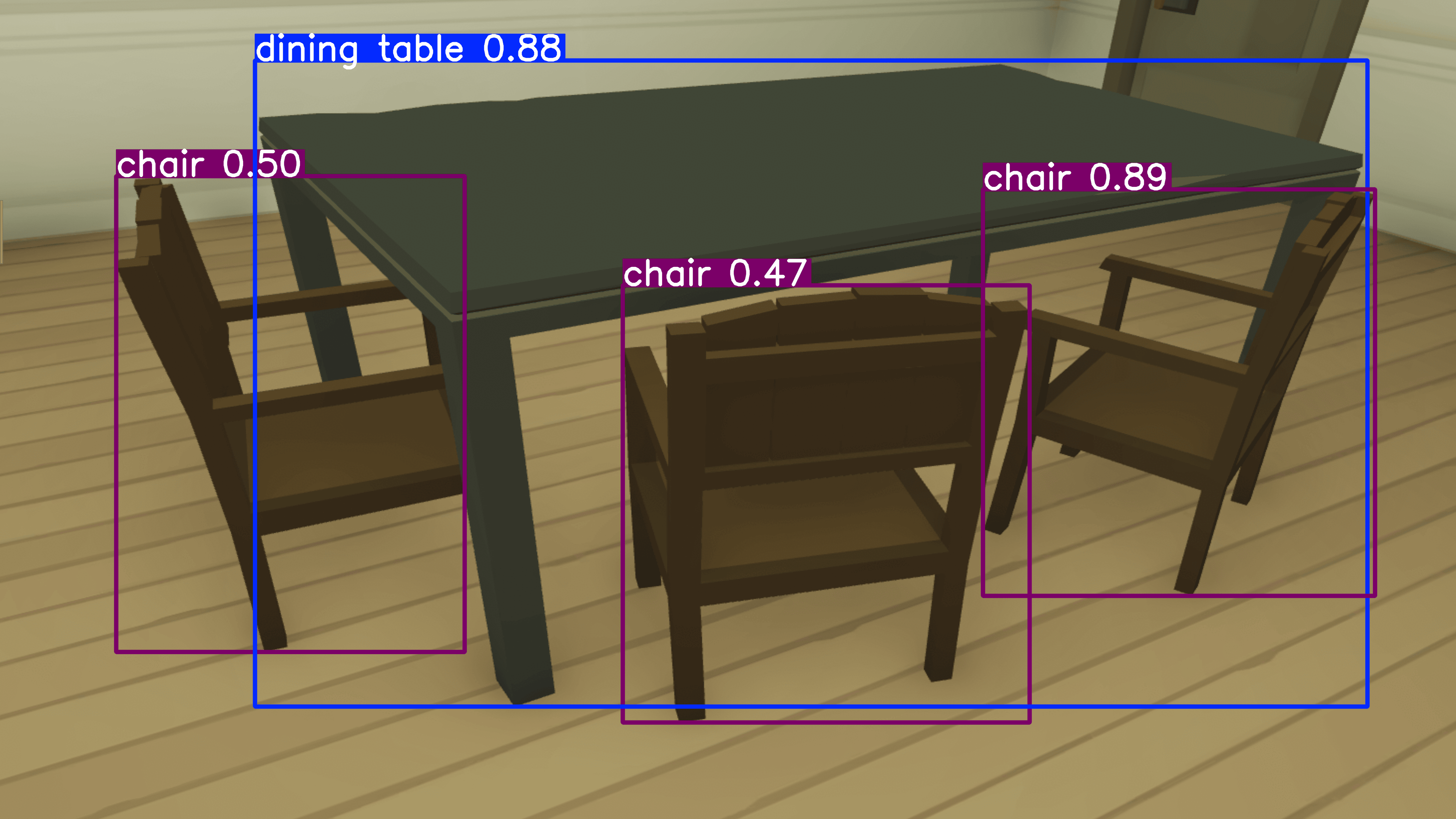}
        \caption{Easy}
    \end{subfigure}
    \hfill
    \begin{subfigure}{0.32\textwidth}
        \includegraphics[width=\linewidth]{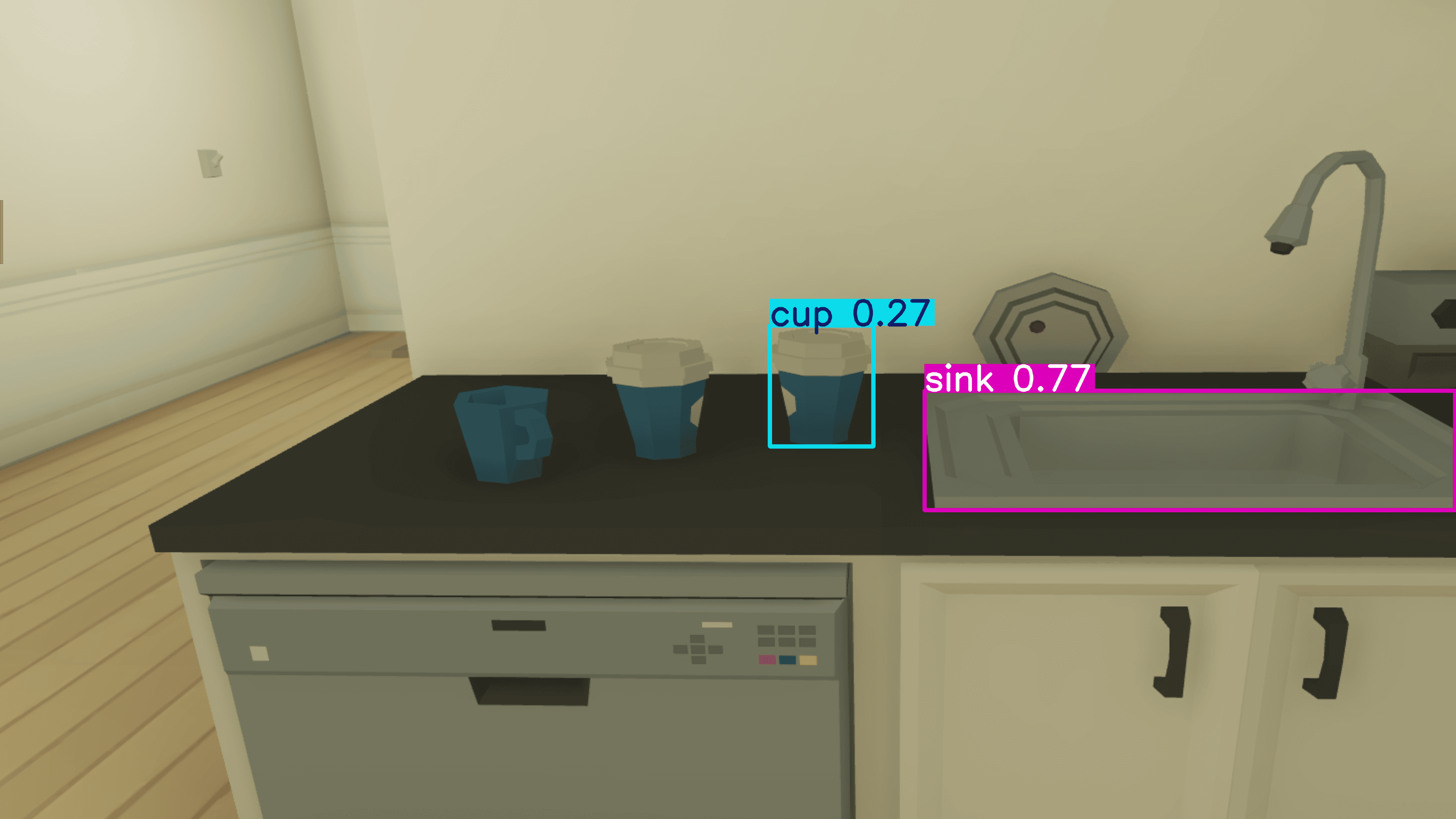}
        \caption{Medium}
    \end{subfigure}
    \hfill
    \begin{subfigure}{0.32\textwidth}
        \includegraphics[width=\linewidth]{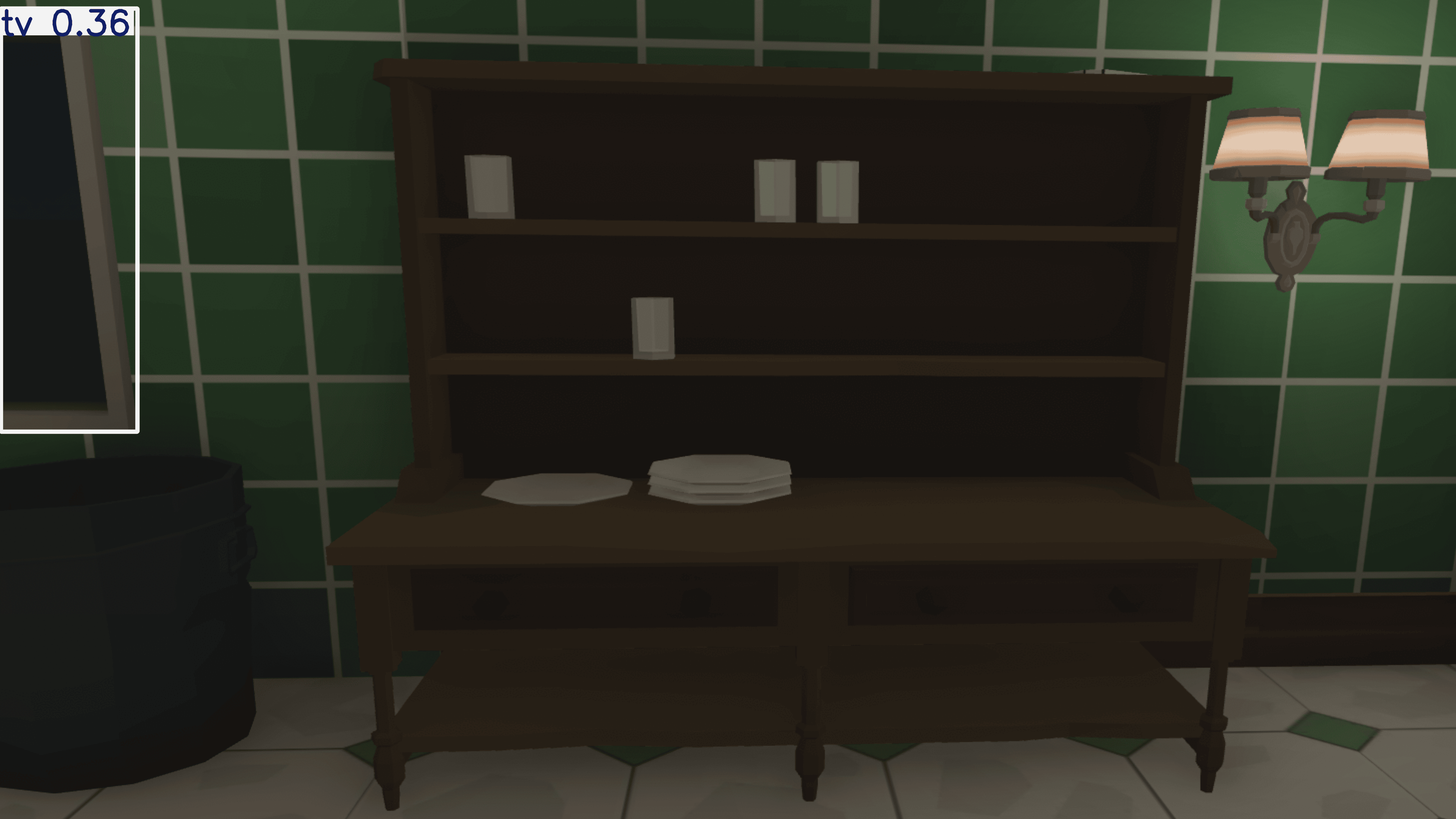}
        \caption{Hard}
    \end{subfigure}
    \caption{Result of YOLOv11}
    \label{fig:yolo_result}
\end{figure}

We observe that GPT-4o struggles to perform entity labeling correctly due to its architecture's inability to process and interpret visual data directly. 
While GPT-4o excels in natural language tasks, it lacks mechanisms for handling spatial and pixel-level information, which are essential for entity labeling. 
To tackle the challenge, we conduct experiments using a task-specific multimodal LLM--Florence-2~\cite{florence2}, which features a specially designed structure~\cite{clip,ViT} and is fine-tuned with image-text data, to label entities across diverse images. As shown in Figure \ref{fig:florence_result}, Florence-2 can successfully label large entities; however, it fails to label small entities, indicating further model tuning is required.
Furthermore, Florence-2 has high computational overheads. To address this, we use a generic object detection deep learning model to draw bounding boxes, then input the images with bounding boxes to GPT-4o for labeling.
As shown in Figure~\ref{fig:yolo_result}, YOLO \cite{yolov11}, a real-time object detection deep learning model, can accurately draw the bounding box of entities, suggesting a possible path for future works. 
Moreover, A recent VR GUI labeling tool named~\cite{li2024groundedguiunderstandingvision} also discusses the inability of existing multimodal LLMs on labeling the heterogeneous GUI elements in VR and design approaches to improve the labeling accuracy.

\subsection{Limitation and Improvement for LLM Reasoning}

\begin{figure}
  \centering
  \includegraphics[width=0.9\linewidth]{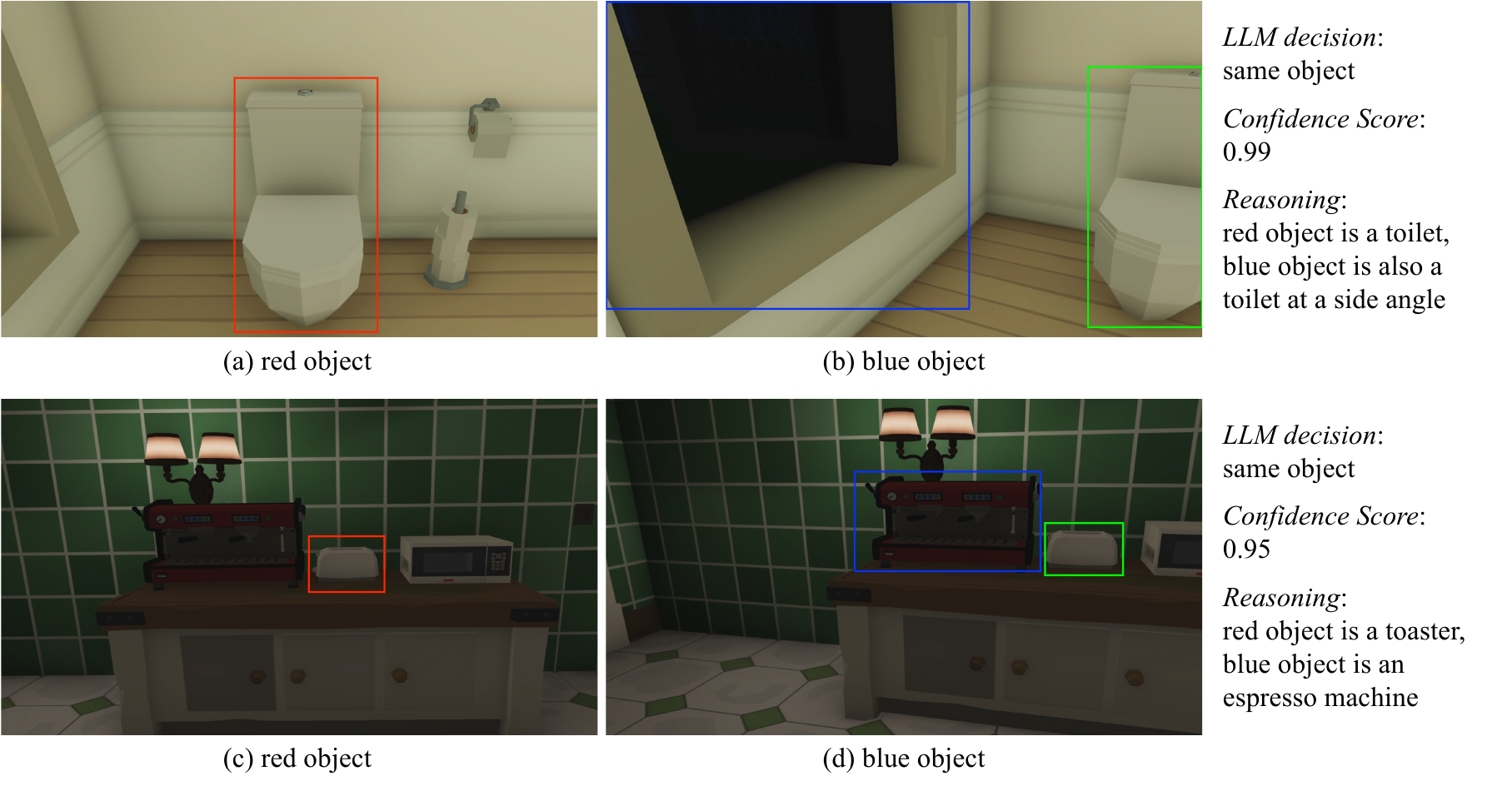}
  \caption{Examples of Unreliable LLM Reasoning}
  \label{fig:discussion_reliability}
\end{figure}

In RQ5, we ask GPT-4o to determine whether two entities across multiple FOVs are identical,
and further report the confidence score and reason for the decision.
We believe a reliable and trustworthy model should be consistent between the explanation and decision. Otherwise, these conflict statements will not be reliable for further automated testing processes.
When examining the reports, we found that the confidence score does not have a notable correlation between the correct and incorrect results. The LLM can be very confident when making mistakes and also when making correct decisions.
This issue is well-known and has already been discussed by many researchers~\cite{xiong2024can, yupengacm2024survey}. A recent study~\cite{yang2024verbalizedconfidencescoresllms} assesses the reliability of confidence scores and finds that it varies by context and is thus unreliable.
Moreover, as shown in Figure~\ref{fig:discussion_reliability}, the reasoning statements may also contain mistakes and conflict with final decisions.
In the first example of comparing the red entity in (a) and the blue entity in (b),
LLM explains that the red entity is a toilet, and the blue entity is a toilet at a side angle. This mistake is caused by incorrectly identifying the labeled entity in green with the entity in blue in the images.
In the second example of comparing the red and blue entities in (c) and (d),
although the LLM successfully identifies that they are different entities, it still makes the decision as identical.

To improve LLM's reliability, we could first adopt AI explainability techniques such as Stable Explanation~\cite{becker2024cyclesthoughtmeasuringllm}, Chain-of-Thought~\cite{wang2023selfconsistency}, and Multiple Answers~\cite{tian-etal-2023-just} to enhance the quality of reasoning. Additionally, we could leverage structured output to create a comprehensive system that compares entities across different features, assigning similarity weights and reporting the determination as a probability score instead of a hard yes-or-no answer. Ultimately, the poorly calibrated confidence scores generated by the contemporary Large Language Model (GPT-4o) underscore fundamental limitations in current architectures, suggesting that targeted fine-tuning may be essential to effectively address this issue.

\subsection{Implications for Future Directions}

\textbf{For VR Developers/Testers.}
RQ2 and RQ3 have demonstrated the potential of leveraging LLMs to identify the test entities for black-box testing in VR exploration. 
VR developers and testers could build lightweight testing tools and frameworks with zero domain knowledge of the context of the tested applications by using LLMs as assistants to provide general knowledge and a reasonable understanding.
These tools can help examine the testing targets, simulate the end-user perspective, and even map the textual testing requirements to the actual entities on the FOVs.
These assistant tools would also assist human testers in quickly identifying potential virtual bugs and accurately applying interactions to verify functional designs. The LLMs could be very helpful for testers with limited technical expertise, helping to further reduce technical debt in VR testing.
Moreover, this black-box approach could be widely applied to all kinds of VR devices and platforms, which can reduce many configuration and compatibility challenges and create many platform-independent testing solutions. 
Additionally, the general knowledge of LLM could potentially be used to design a general virtual assistant in the Human-Computer Interaction (HCI) community to improve the quality and accessibility of VR applications.

\textbf{For SE Researchers.}
While demonstrating LLM potential for VR entity identification and scene understanding (RQ1-RQ3), this study also reveals critical limitations, particularly in visual grounding for entity labeling (RQ4) and reasoning reliability for cross-FOV entity determination (RQ5), creating significant opportunities for SE research. Future work could focus on advancing multimodal models beyond general-purpose architectures like GPT-4o, potentially through fine-tuning on VR-specific datasets, exploring specialized architectures (e.g., Florence-2-inspired), or developing hybrid frameworks combining robust visual detectors (e.g., YOLO) with LLM reasoning. Investigating few-shot learning for rapid adaptation to new VR applications is also crucial. Furthermore, the unreliable confidence scores and reasoning observed (RQ5) necessitate research into improving model calibration and trustworthiness for VR tasks, applying tailored explainability techniques (e.g., CoT, Stable Explanation), and designing systems with probabilistic outputs. Ultimately, this study suggests the development of advanced automated VR testing frameworks, encompassing more sophisticated test oracles capable of verifying complex interactions and states, and underscores the continued need for large-scale, annotated VR datasets to drive progress in this research area.

\section{Threats to Validity} \label{sec:thread}

On internal validity, the first threat is the potential bias from a single LLM, GPT-4o.
By the time we conduct the study, GPT-4o is the most balanced and advanced model for handling different types of tasks.
So, we select this model as the subject of the study. Moreover, our study focuses on evaluating the potential instead of finding the best solution for all the study tasks. We do not intend to compare the results of different LLMs, but mainly report the baseline of the accuracy rate.
The second threat is the potential bias from groundtruth labeling in RQ2 and RQ3
since there might exist multiple ways to describe the feature and the entire scene.
For example, a room with toys and cribs can be described as a "nursery room" and also a "children's room" since both indicate a space designed for children. However, there are subtle differences, as the first term often refers to infants and toddlers, while the latter refers to children of any age.
We use the open coding method~\cite{charmaz2006constructing} to avoid such potential bias.
Specifically, before generating the results from LLM, we ask two annotators to label the screenshots independently, and then we let the third annotator determine the consistency and make the final decision.
Ultimately, we ask three annotators to vote on accepting the LLM results with the groundtruths.

The third threat is the potential bias in collecting feature candidates for RQ2. 
A real-world entity could have numerous features covering how it looks, what it does, and how it interacts with users and its environment. In this study, we focus on VR entity identification and navigation during the GUI exploration. As a result, visual features are the most important compared to other features. To capture as many visual features as possible, we collect $9$ features from $17$ relevant studies in the field of Cognitive Science and Computer Vision.

Regarding external validity, the main concern is the representativeness of the subjects studied.
Among all the VR apps, we select a highly rated cartoon-like simulation game with $87.8\%$ positive reviews. We do not pick large-scale real-world simulation VR apps due to the difficulty of entity controls. Large-scale VR apps typically display thousands of entities on a single screen, making it impractical for the LLM to label every entity and making it impossible to manually label the groundtruths.
Since our goal is to establish a baseline for LLM performance and demonstrate its potential, we begin with simplified scenes that, while minimal, still include sufficient detail beyond basic Greyboxing shapes to support meaningful evaluation of the designed tasks.
In total, we obtain $270$ screenshots from various rooms and viewpoints within a virtual house, labeling $216$ distinct entities ranging from a large sofa in the living room to a small mug on the countertop. Furthermore, each scene is built under three various lighting conditions and five different view angles.
We believe this relatively comprehensive dataset provides a reasonable foundation for assessing the general performance of the LLM in exploration testing.
\section{Related Work} \label{sec:related}

\subsection{Large Language Model on GUI Automation}
A vast amount of GUI automation approaches~\cite{zhang2024largelanguagemodelbrainedgui} have been proposed since the advent of Large Language Models (LLMs). This is due to the ability of LLMs to understand natural language, generate code, and process visual inputs, which makes GUIs of real-world applications more explainable and feasible for performing different types of tasks.
Current applications of automation focus on several areas,
including general testing~\cite{GPTDroid, AUITestAgent, AXNav}, where natural language tasks are interpreted and executed programmatically, and also special testing with tasks of text input generation~\cite{QTypist}, bug replay~\cite{CrashTranslator}.
Moreover, LLMs are increasingly used to build virtual assistants that enable more complex, context-aware interactions with device GUIs, supporting both textual and voice commands~\cite{GPTVoiceTasker, EasyAsk, PromptRPA}.
These approaches are applicable across a broad spectrum of platforms, from web and mobile applications to desktop environments.
However, little progress has been made in Virtual Reality applications.
Qin et al.~\cite{qin2024utilizing} offered an initial investigation of GenAI (GPT-4) for VR exploration tasks via FOV analysis. This preliminary, limited-scope study focused on object identification (achieving 63\% accuracy), organization, and action suggestion, noting difficulties particularly with organization and localization.
The current LLM practices on GUI demonstrate the possibility of building an automated AI agent for VR exploration testing.

\subsection{Automated Testing in VR and 3D Game}
The idea of automated testing in VR was proposed two decades ago~\cite{Cruz-Neira}.
Many automated user interaction evaluation approaches~\cite{Patrick, karakaya2022automated} were designed in the CHI community.
In Software Testing, recent works that focus on GUI include scene inspection~\cite{vrtest, vrguide}, where researchers try to effectively find the best angle of view that contains as many test entities as possible; GUI testing code study~\cite{rzig2023vr_testing}, where researchers reviewed hundreds of GUI-related test methods from Unity VR applications; Automated Biomechanical Testing~\cite{SIM2VR}, where researchers used a biomechanical model to simulate the user's muscle actions for test input.
Moreover, exploration testing using Reinforcement Learning agents~\cite{Angelo2023ASE, Joseph2022ATEST} has been proposed in 3D game testing.
Unfortunately, many of these exploration techniques are based on white-box testing or require extra effort to adopt the designed framework. With over six major VR brands and over five VR engines on the market, it is a challenge to apply these techniques to support most VR apps.  
Recently, a VR-based automated GUI element recognition tool named Orienter~\cite{li2024groundedguiunderstandingvision} has been proposed.
In this paper, the authors focus on identifying the objects and drawing the bounding boxes around them through the feedback prompt engineering, and these boxes could be later used to reduce the action space in testing.
The Orienter can be used to address the entity labeling limitations we observed in RQ4. However, this work will not be able to support the other testing tasks, such as feature identification, scene and spatial analysis, and entity determination through multiple FOVs from Studies 2, 3, and 5.

\subsection{Object Detection and Labeling}
Ren et al. \cite{FRCNN} propose the Region Proposal Network, which generates region proposals efficiently within the model itself. Faster R-CNN becomes one of the foundational models, significantly improving the speed and accuracy of object detection.
Liang et al. \cite{RCNN} proposed the CBNet, which constructs detectors by integrating multiple identical backbones with composite connections. 
For precise detection, Carion et al. \cite{DETR} introduce the encoder-decoder architecture to build the DETR model. 
Zhang et al. \cite{DINO, dndetr, dabdetr} present DINO, an enhanced version of DETR, to introduce a denoising training mechanism that uses noisy anchor boxes as additional supervision during training.
In real-time detection, Redmon et al. \cite{YOLO} present the YOLO model, which frames it as a regression problem, predicting bounding boxes and class probabilities directly from full images. 
In video object detection, Wang et al. \cite{VisTR} present VisTR, which views it as a parallel sequence decoding/prediction problem. 
Some vision task-specific LLMs could also handle the object detection task, Yuan et al. \cite{florence} present Florence, a unified computer vision foundation model, which pretrains on large-scale curated data, has exceptional
zero-shot and few-shot capabilities.

\section{Conclusion} \label{sec:conclusion}
In this paper, we construct a large-scale dataset and conduct a case study to explore the potential of LLMs in assisting VR application exploration testing through field of view (FOV) analysis. As a result, the GPT-4o achieves an average accuracy rate of $71.30\%$ in entity detection using a chain-of-thought prompt design. We then adopt this prompt design to identify the core features of entities, achieving an accuracy rate of at least $94.8\%$.
The LLM also demonstrates good potential for scene recognition and spatial understanding, but performs poorly in labeling the detected entities in images.
Moreover, the combination of core features—color, shape, and placement—achieves the highest accuracy in determining identical entities across multiple FOVs.
Ultimately, we believe the LLM has the potential to assist in GUI automation testing, but it also has limitations that need to be addressed in future work.

\bibliography{sn-bibliography} 

\end{document}